    	\catcode`\"=12
	\font\black=cmbx10
\font\sblack=cmbx7
\font\ssblack=cmbx5
\font\blackital=cmmib10  \skewchar\blackital='177
\font\sblackital=cmmib7  \skewchar\sblackital='177
\font\ssblackital=cmmib5  \skewchar\ssblackital='177
\font\sanss=cmss10
\font\ssanss=cmss8 scaled 900
\font\sssanss=cmss8 scaled 600
\font\blackboard=msbm10
\font\sblackboard=msbm7
\font\ssblackboard=msbm5
\font\caligr=zplmr7y
\font\scaligr=zplmr7y scaled 650
\font\sscaligr=zplmr7y scaled 500

\font\fraktur=eufm10
\font\sfraktur=eufm7
\font\ssfraktur=eufm5

\font\amsa=msam10

\font\syx=cmsy10 scaled 1300

\font\sya=cmsy10 scaled 650
    
    \def\all#1{\setbox0=\hbox{\lower1.5pt\hbox{\bsymb
       \char"38}}\setbox1=\hbox{$_{#1}$} \box0\lower2pt\box1\;}
    \def\exi#1{\setbox0=\hbox{\lower1.5pt\hbox{\bsymb \char"39}}
       \setbox1=\hbox{$_{#1}$} \box0\lower2pt\box1\;}

\def\PR{{\setbox0=\hbox{\syx\char"02}\box0}}

\def\RP{{\setbox0=\hbox{\sya\char"7B}\box0}}

\font\bsymb=cmsy10 scaled\magstep2
\font\CL=cmcsc10

\def\tx#1{{\fam0\relax#1}}

\newfam\bifam
\textfont\bifam=\blackital
\scriptfont\bifam=\sblackital
\scriptscriptfont\bifam=\ssblackital

\newfam\blfam
\textfont\blfam=\black
\scriptfont\blfam=\sblack
\scriptscriptfont\blfam=\ssblack
\def\rbl#1{{\fam\blfam\relax#1}}

\newfam\bbfam
\textfont\bbfam=\blackboard
\scriptfont\bbfam=\sblackboard
\scriptscriptfont\bbfam=\ssblackboard
\def\bb#1{{\fam\bbfam\relax#1}}

\newfam\ssfam
\textfont\ssfam=\sanss
\scriptfont\ssfam=\ssanss
\scriptscriptfont\ssfam=\sssanss
\def\ss#1{{\fam\ssfam\relax#1}}

\newfam\clfam
\textfont\clfam=\caligr
\scriptfont\clfam=\scaligr
\scriptscriptfont\clfam=\sscaligr

\newfam\frfam
\textfont\frfam=\fraktur
\scriptfont\frfam=\sfraktur
\scriptscriptfont\frfam=\ssfraktur

\def\hpb#1{\setbox0=\hbox{${#1}$}
    \copy0 \kern-\wd0 \kern.2pt \box0}
\def\vpb#1{\setbox0=\hbox{${#1}$}
    \copy0 \kern-\wd0 \raise.08pt \box0}
\def\pmb#1{\setbox0\hbox{${#1}$} \copy0 \kern-\wd0 \kern.2pt \box0}
\def\pmbb#1{\setbox0\hbox{${#1}$} \copy0 \kern-\wd0
      \kern.2pt \copy0 \kern-\wd0 \kern.2pt \box0}
\def\pmbbb#1{\setbox0\hbox{${#1}$} \copy0 \kern-\wd0
      \kern.2pt \copy0 \kern-\wd0 \kern.2pt
    \copy0 \kern-\wd0 \kern.2pt \box0}
\def\pmxb#1{\setbox0\hbox{${#1}$} \copy0 \kern-\wd0
      \kern.2pt \copy0 \kern-\wd0 \kern.2pt
      \copy0 \kern-\wd0 \kern.2pt \copy0 \kern-\wd0 \kern.2pt \box0}
\def\pmxbb#1{\setbox0\hbox{${#1}$} \copy0 \kern-\wd0 \kern.2pt
      \copy0 \kern-\wd0 \kern.2pt
      \copy0 \kern-\wd0 \kern.2pt \copy0 \kern-\wd0 \kern.2pt
      \copy0 \kern-\wd0 \kern.2pt \box0}

\def\blacktriangle{{\setbox0=\hbox{\amsa\char"4E}\box0}} 
\def\leqslant{{\setbox0=\hbox{\amsa\char"36}\box0}} 
\def\geqslant{{\setbox0=\hbox{\amsa\char"3E}\box0}}

\mathchardef\za="710B  
\mathchardef\zb="710C  
\mathchardef\zg="710D  
\mathchardef\zd="710E  
\mathchardef\zve="710F 
\mathchardef\zz="7110  
\mathchardef\zh="7111  
\mathchardef\zvy="7112 
\mathchardef\zi="7113  
\mathchardef\zk="7114  
\mathchardef\zl="7115  
\mathchardef\zm="7116  
\mathchardef\zn="7117  
\mathchardef\zx="7118  
\mathchardef\zp="7119  
\mathchardef\zr="711A  
\mathchardef\zs="711B  
\mathchardef\zt="711C  
\mathchardef\zu="711D  
\mathchardef\zvf="711E 
\mathchardef\zq="711F  
\mathchardef\zc="7120  
\mathchardef\zw="7121  
\mathchardef\ze="7122  
\mathchardef\zy="7123  
\mathchardef\zvp="7124 
\mathchardef\zvr="7125 
\mathchardef\zvs="7126 
\mathchardef\zf="7127  
\mathchardef\zG="7000  
\mathchardef\zD="7001  
\mathchardef\zY="7002  
\mathchardef\zL="7003  
\mathchardef\zX="7004  
\mathchardef\zP="7005  
\mathchardef\zS="7006  
\mathchardef\zU="7007  
\mathchardef\zF="7008  
\mathchardef\zC="7009  
\mathchardef\zW="700A  

    \font\kropa=lcircle10 scaled 1700
    \def\tdot{\setbox0=\hbox{\kropa \char"70} \kern1.5pt \raise.35pt \box0}
    \def\sdot{\setbox0=\hbox{\kropa \char"70} \kern1.5pt \raise2.9pt \box0}

    \def\N{{\bb N}}
    \def\R{{\bb R}}

    \def\sH{{\ss H}}

    \def\sT{{\ss T}}
    
    \def\sV{{\ss V}}

    \def\st{{\ss t}}

    \def\rD{{\tx D}}
    \def\rd{{\tx d}}
    \def\ri{{\tx i}}
   \def\wU{{\widetilde U}}

    \def\w#1{{\widetilde #1}}

	\catcode`\"=\active

	\newcount\secnum  \secnum=0 
	\newcount\subsecnum
	\newcount\subsubsecnum 

	\newcount\Asecnum  \secnum=0 
	\newcount\Asubsecnum

	\newcount\tcount  \tcount=0
	\newcount\pcount  \pcount=0
	\newcount\dcount  \dcount=0
	\newcount\ecount  \ecount=0

	\def\Theorem#1{\global\advance\tcount by 1
			\vskip2pt
			\noindent{\CL Theorem \the\tcount.} {\it #1}\par\vskip2pt}

	\def\Proposition#1{\global\advance\pcount by 1
			\vskip2pt
			\noindent{\CL Proposition \the\pcount.} {\it #1}\par\vskip2pt}

	\def\Definition#1{\global\advance\dcount by 1
			\vskip2pt
			\noindent{\CL Definition \the\dcount.} #1
			{\hfill \blacktriangle}\par\vskip2pt}

	\def\Example#1{\global\advance\ecount by 1
			\vskip2pt
			\noindent{\CL Example \the\ecount.} #1
			{\hfill \blacktriangle}\par\vskip2pt}

	\def\Proof#1{\noindent{\CL Proof:} #1{\hfill \vrule height 3pt width 5pt depth 2pt}\par\vskip2pt}

	\def\sect#1{ 
            \global\advance \secnum by 1 \subsecnum=0 \subsubsecnum=0 
			\vskip2pt
            \noindent{\bf \the\secnum.\ #1 \vskip-4.5mm}
			\nobreak
			\leftline{}}

	\def\ssect#1{
	    	\global\advance\subsecnum by 1 \subsubsecnum=0
			\vskip1pt
            \noindent{\bf \the \secnum.\the\subsecnum.\ #1 \vskip-4.5mm}
		    \nobreak
			\leftline{}}

	\def\sssect#1{\global\advance\subsubsecnum by 1 
            \vskip1pt
			\noindent{\bf \the\secnum.\the\subsecnum.\the\subsubsecnum.\ #1 \vskip-4.5mm} 
	        \nobreak
		    \leftline{}}

	\def\Asect#1{ 
            \global\advance \Asecnum by 1 \Asubsecnum=0 
			\vskip2pt
            \noindent{\bf \the \Asecnum.\ #1 \vskip-4.5mm}
			\nobreak
			\leftline{}}

	\def\Assect#1{
	    	\global\advance\Asubsecnum by 1 
			\vskip1pt
            \noindent{\bf A\the \Asecnum .\the\Asubsecnum\  #1 \vskip-4.5mm}
		    \nobreak
			\leftline{}}

	\font\Tfont=cmb10 scaled 1300

	\font\tfont=cmb10 

	\def\Title#1{\vskip2mm\centerline{\Tfont #1}\vskip1.5mm}

	\def\title#1{\vskip1mm\leftline{\tfont #1}}

	\def\compose#1#2#3#4#5#6{{\setbox0=\hbox{\raise#2\hbox{\kern#3\hbox{${#1}$}}}
\setbox1=\hbox{\raise#5\hbox{\kern#6\hbox{${#4}$}}}\box0\box1}}

	\def\position#1#2#3{{\setbox0=\hbox{\raise#2\hbox{\kern#3\hbox{${#1}$}}}\box0}}

	\def\Sim#1{\kern.5pt\setbox0=\hbox{$\VPU{8pt}^{\scriptscriptstyle{#1}}$}\setbox1=\hbox{$\sim$}
			\setbox2=\hbox{\kern.5\wd0\copy1\kern-.5\wd0\kern-3pt\copy0}\box2\kern2.7pt}

	\def\RF{\null}
	\def\HEQ#1{{\vbox{\hrule width#1  \vskip1.8pt \hrule width#1}}}
	
	\def\VPU#1{{\vrule height#1 width0pt depth0pt}}
	\def\VPD#1{{\vrule height0pt width0pt depth#1}}

	\def\lpr{{\setbox0=\hbox{\vrule height .15pt width 3.5pt depth 0pt}
\setbox1=\hbox{\vrule height 5.8pt width .3pt depth 0pt}\kern2pt\box0\box1\kern3pt}}

	\def\rpr{{\setbox0=\hbox{\vrule height .15pt width 3.5pt depth 0pt}
\setbox1=\hbox{\vrule height 5.8pt width .3pt depth 0pt}\kern3pt\box0\kern-3.5pt\box1\kern6.5pt}}

	\def\fpr#1{\kern-4pt\setbox0=\hbox{$\VPD{5pt}_{#1}$}\setbox1=\hbox{$\times$}
			\setbox2=\hbox{\kern.5\wd0\copy1\kern-.5\wd0\kern-3pt\copy0}\box2\kern0pt}

	\def\List{\parindent=30pt}
	\def\endList{\parindent=20pt}
	\def\dacapo{\vskip0pt\indent}

	\def\*{{\textstyle *}}

	\def\polar{{\textstyle\circ}}

\def\ssT{{\scriptscriptstyle {\ss T}}}

\def\*{{\textstyle *}}
\def\s*{{\scriptstyle *}}
\def\srP{{\position {\RP}{3pt}{.6pt}}}

	\def\leqs{\kern3pt\leqslant\kern3pt}
  	\def\geqs{\geqslant}

    \hsize=37pc
    \hoffset=-10pt
    \vsize=52pc
    \voffset=6pt
    \input xy
    \xyoption{all}

        \input pictex
        \input DCpic.sty



\def\gr{{\rm gr}}

\def\im{{\rm im}}

\def\osT{\sT \position{\scriptscriptstyle\circ}{9pt}{-5.5pt}\kern3pt}

\def\WMT{
		\centerline{W\l odzimierz M. Tulczyjew }
		\centerline{Valle San Benedetto, 2 }
        \centerline{62030 Monte Cavallo, Italy }
		\centerline{Associated with }
        \centerline{Division of Mathematical Methods in Physics}
        \centerline{University of Warsaw}
        \centerline{Ho\.{z}a 74, 00-682 Warszawa}
	\centerline{and}
        \centerline{Istituto Nazionale di Fisica Nucleare,}
        \centerline{Sezione di Napoli}
        \centerline{Complesso Universitario di Monte Sant'Angelo}
        \centerline{Via Cinthia, 80126 Napoli, Italy}
		\centerline{{\tt tulczy@libero.it} }
		\vskip1mm
		}

\def\PUR{
        \centerline{Pawe\l\ Urba\'nski}
        \centerline{Division of Mathematical Methods in Physics}
        \centerline{University of Warsaw}
        \centerline{Ho\.{z}a 74, 00-682 Warszawa}
        \centerline{{\tt urbanski@fuw.edu.pl}}
		\vskip1mm
		}

\catcode`\"=12

\catcode`\"=\active

\def\text#1{{\rm\hskip2mm #1 \hskip2mm}}



    \Title{Liouville structures}
\WMT
    \vskip2mm
\PUR

    \catcode`\"=12
    \font\kropa=lcircle10 scaled 2000
    \def\rbl{\setbox0=\hbox{\kropa \char"70} \kern2.1pt \raise2.5pt \box0 \kern.9pt}
    \catcode`\"=\active

    \def\shpb#1{\setbox0=\hbox{${#1}$}\copy0 \kern-\wd0 \kern.2pt \box0}
    \def\svpb#1{\setbox0=\hbox{${#1}$}\copy0 \kern-\wd0 \raise.19pt \box0}
    \def\cmb#1{\shpb{\svpb{#1}}}
    \def\cmbb#1{\cmb{\cmb{#1}}}
    \def\cmbbb#1{\cmb{\cmbb{#1}}}
    \def\tplus{\,\cmbbb+\,}
    \def\tdot{\rbl}

        \sect{Introduction.}
    A {\it Liouville structure} is a structure isomorphic to a cotangent vector fibration.  A Liouville structure is an
essential ingredient of every variational formulation of a physical theory.  For reasons of interpretation the Liouville
structure can not be replaced by the corresponding cotangent fibration.  We give a precise definition of Liouville
structures, study their properties, and give examples used in variational formulations of mechanics.

        \sect{Tangent and cotangent fibrations.}
    Let $Q$ be a differential manifold.  In the set $C^\infty(Q|\R)$ of differentiable curves in $Q$ we introduce an
equivalence relation.  Two curves $\zg$ and $\zg'$ are equivalent if $\zg'(0) = \zg(0)$ and $\rD (f \circ \zg')(0) = \rD (f
\circ \zg)(0)$ for each differentiable function $f \,\colon Q \rightarrow \R$.  Equivalence classes are called {\it tangent
vectors}.  The set of all tangent vectors will be denoted by $\sT Q$.  The equivalence class of a curve $\zg \,\colon \R
\rightarrow Q$ will be denoted by $\st\,\zg(0)$.  We have the surjective mapping
        $$\zt_Q \,\colon \sT Q \rightarrow Q \,\colon \st\,\zg(0) \mapsto \zg(0)
                                                                                                                \eqno(1)$$
    called the {\it tangent fibration}.

    Let $C^\infty(\R|Q)$ denote the algebra of differentiable functions on a differential manifold $Q$.  In the set
$C^\infty(\R|Q) \times Q$ we introduce an equivalence relation.  Two pairs $(f,q)$ and $(f',q')$ are equivalent if $q' = q$
and
        $$\rD(f' \circ \zg)(0) = \rD(f \circ \zg)(0)
                                                                                                                \eqno(2)$$
    for each differentiable curve $\zg \,\colon \R \rightarrow Q$ such that $\zg(0) = q$.  The set of equivalence classes
denoted by $\sT^\*Q$ is called the {\it cotangent bundle} of $Q$.  Elements of $\sT^\*Q$ are called {\it cotangent
vectors}.  The equivalence class of $(f,q)$ denoted by $\rd f(q)$ is called the {\it differential} of $f$ at $q$.  The
mapping
        $$\zp_Q \,\colon \sT^\*Q \rightarrow Q \,\colon \rd f(q) \mapsto q
                                                                                                                \eqno(3)$$
    is called the {\it cotangent bundle projection}.  Linear operations in fibres of the cotangent fibration have natural
definitions
        $$+ \,\colon \sT^\* Q \fpr{(\zp_Q,\zp_Q)} \sT^\* Q \rightarrow \sT^\* Q \,\colon (\rd f_1(q),\rd f_2(q)) \mapsto
\rd(f_1 + f_2)(q)
                                                                                                                \eqno(4)$$
    and
        $$\cdot\, \,\colon \R \times \sT^\* Q \rightarrow \sT^\* Q \,\colon (k,\rd f(q)) \mapsto \rd(kf)(q).
                                                                                                                \eqno(5)$$

    The mapping
        $$\langle \,\; ,\; \rangle\VPD{4pt}_Q \,\colon \sT^\*Q \fpr{(\zp_Q,\zt_Q)} \sT Q \rightarrow \R \,\colon (\rd
f(a),\st\,\zg(0)) \mapsto \rD(f \circ \zg)(0)
                                                                                                                \eqno(6)$$
    is a differentiable nondegenerate pairing linear in its covector argument.  There is a unique structure of a vector
fibration for $\zt_Q$ such that the canonical pairing \RF(6) is bilinear.  The tangent fibration and the cotangent
fibration are a dual pair of vector fibrations.

        \sect{The tangent of a vector fibration.}
    Let    
    \vskip1mm
        $$\vcenter{
        \begindc{0}[1]
        \obj(000,45)[01]{$P$}
        \obj(000,00)[00]{$Q$}
        \mor{01}{00}[8,8]{$\zp$}[-1,0]
        \enddc}
                                                                                                                \eqno(7)$$
    \vskip2mm
    \noindent be a vector fibration with operations
        $$+\, \,\colon P \fpr{(\zp,\zp)} P \rightarrow P
                                                                                                                \eqno(8)$$
    and
        $$\cdot\, \,\colon \R \times P \rightarrow P.
                                                                                                                \eqno(9)$$
    Let
        $$O_\zp \,\colon M \rightarrow P
                                                                                                                \eqno(10)$$
    be the zero section.

    The tangent fibration
    \vskip1mm
        $$\vcenter{
        \begindc{0}[1]
        \obj(000,45)[01]{$\sT P$}
        \obj(000,00)[00]{$P$}
        \mor{01}{00}[8,8]{$\zt_P$}[-1,0]
        \enddc}
                                                                                                                \eqno(11)$$
    \vskip2mm
    \noindent is a vector fibration with operations
        $$+\, \,\colon \sT P \fpr{(\zt_P,\zt_P)} \sT P \rightarrow \sT P
                                                                                                                \eqno(12)$$
    and
        $$\cdot\, \,\colon \R \times \sT P \rightarrow \sT P,
                                                                                                                \eqno(13)$$
    and the zero section
        $$O_{\zt_P} \,\colon P \rightarrow \sT P.
                                                                                                                \eqno(14)$$

    The diagram
    \vskip1mm
        $$\vcenter{
        \begindc{0}[1]
        \obj(000,45)[01]{$\sT P$}
        \obj(000,00)[00]{$\sT Q$}
        \mor{01}{00}[8,8]{$\sT\zp$}[-1,0]
        \enddc}
                                                                                                                \eqno(15)$$
    \vskip2mm
    \noindent is again a vector fibration with operations
        $$\tplus \,\colon \sT P \fpr{(\sT\zp,\sT\zp)} \sT P \rightarrow \sT P \,\colon (\st\,\zz_1(0),\st\,\zz_2(0)) \mapsto
\st\,(\zz_1 + \zz_2)(0)
                                                                                                                \eqno(16)$$
    and
        $$\tdot\, \,\colon \R \times \sT P \rightarrow \sT P \,\colon (k,\st\,\zz(0)) \mapsto \st\,(k\zz)(0),
                                                                                                                \eqno(17)$$
    and the zero section
        $$O_{\sT\zp} = \sT O_\zp \,\colon \sT M \rightarrow \sT P.
                                                                                                                \eqno(18)$$

    In the definition of the operation $\tplus$ the curves $\zz_1 \,\colon \R \rightarrow P$ and $\zz_1 \,\colon \R
\rightarrow P$ representing the vectors $\st\,\zz_1(0)$ and $\st\,\zz_2(0)$ are chosen to satisfy the condition $\zp \circ
\zz_2 = \zp \circ \zz_1$.

    There are alternative constructions of the operations $\tplus$ and $\tdot$.  The operation $\tplus$ is obtained from
the tangent mapping
        $$\sT+ \,\colon \sT(P \fpr{(\zp,\zp)} P) \rightarrow \sT P
                                                                                                                \eqno(19)$$
    by identifying the space $\sT(P \fpr{(\zp,\zp)} P)$ with $\sT P \fpr{(\sT\zp,\sT\zp)} \sT P$.  The operation $\tdot$ is
constructed from the tangent mapping
        $$\sT\cdot\, \,\colon \sT(\R \times P) \rightarrow \sT P.
                                                                                                                \eqno(20)$$
    The space $\sT(\R \times P)$ is identified with $\sT\,\R \times \sT P = \R^2 \times \sT P$ and then the operation
$\tdot$ is introduced as the mapping
        $$\tdot\, \,\colon \R \times \sT P \rightarrow \sT P \,\colon (k,w) \mapsto (k,0)\, \sT\!\cdot \,w.
                                                                                                                \eqno(21)$$

    The diagram
    \vskip1mm
        $$\vcenter{\xymatrix@R+3mm @C+10mm{{\sT P} \ar[d]_*{\zt_P} \ar[r]^*{\sT\zp} &
            \sT Q \ar[d]_*{\zt_Q} \cr
            P \ar[r]^*{\zp} & Q}}
                                                                                                                \eqno(22)$$
    \vskip2mm
    \noindent is a vector fibration morphism.  We show that also the diagram
    \vskip1mm
        $$\vcenter{\xymatrix@R+3mm @C+10mm{{\sT P} \ar[r]^*{\zt_P} \ar[d]_*{\sT\zp} &
            P \ar[d]_*{\zp} \cr
            \sT Q \ar[r]^*{\zt_Q} & Q}}
                                                                                                                \eqno(23)$$
    \vskip2mm
    \noindent is a vector fibration morphism.  If  vectors $w_1 \in \sT P$ and $w_2 \in \sT P$ such that $\sT\zp(w_2) =
\sT\zp(w_1)$ are represented by curves $\zz_1 \,\colon \R \rightarrow P$ and $\zz_1 \,\colon \R \rightarrow P$ chosen to
satisfy the condition $\zp \circ \zz_2 = \zp \circ \zz_1$, then

        $$\eqalign{
    \zt_P(w_1 \tplus w_2) &= \zt_P(\st\,\zz_1(0) \tplus \st\,\zz_2(0)) \cr
            &= \zt_P(\st\,(\zz_1 + \zz_2))(0) \cr
            &= (\zz_1 + \zz_2)(0) \cr
            &= \zz_1(0) + \zz_2(0) \cr
            &= \zt_P(w_1) + \zt_P(w_2).}
                                                                                                                \eqno(24)$$
    Similarily, if $w \in \sT P$ is represented by a curve $\zz \,\colon \R \rightarrow P$, then
        $$\eqalign{
    \zt_P(k \tdot w) =& \zt_P(k \tdot \st\,\zz(0)) \cr
            &= \zt_P(\st\,(k\zz)(0)) \cr
            &= (k\zz)(0) \cr
            &= k\zz(0) \cr
            &= k\zt_P(w).}
                                                                                                                \eqno(25)$$
    This completes the proof.

    The space $\sT P$ with its two vector bundle structures forms a {\it double vector fibration} best represented by
the diagram
    \vskip1mm
        $$\vcenter{\xymatrix@R+3mm @C+3mm{&{\sT P} \ar[dl]_*{\zt_P} \ar[dr]^*{\sT\zp} & \cr
            P \ar[dr]^*{\zp} && \sT Q \ar[dl]_*{\zt_Q} \cr
                    & Q &}}
                                                                                                                \eqno(26)$$
    \vskip2mm
    \noindent We use the concept of a double vector fibration and structures related to double vector fibrations as introduced in [8],
[3], [5], and [2]. The space
        $$C = \left\{w \in \sT P ;\; \sT\zp(w) = O_{\zt_Q}(\zt_Q(\sT\zp(w)), \zt_P(w) = O_\zp(\zp(\zt_P(w)) \right\}
                                                                                                                \eqno(27)$$
    is the {\it core of the double vector fibration} and the diagram
    \vskip1mm
        $$\vcenter{
        \begindc{0}[1]
        \obj(000,45)[01]{$C$}
        \obj(000,00)[00]{$Q$}
        \mor{01}{00}[8,8]{$\ze$}[-1,0]
        \enddc}
                                                                                                                \eqno(28)$$
    \vskip2mm
    \noindent with
        $$\ze \,\colon C \rightarrow Q \,\colon w \mapsto \zp(\zt_P(w)) = \zt_Q(\sT\zp(w)
                                                                                                                \eqno(29)$$
    is the {\it core fibration}.  The core fibration is a vector fibration.  The operation
        $$+\, \,\colon C \fpr{(\ze,\ze)} C \rightarrow C
                                                                                                                \eqno(30)$$
    is derived either from the operation \RF(12) or from \RF(16).  If $w_1$ and $w_2$ are elements of the core such that
$\ze(w_2) = \ze(w_1)$, then $\zt_P(w_2) = \zt_P(w_1)$ and $\sT\zp(w_2) = \sT\zp(w_1)$.  Both operations \RF(12) and \RF(16)
can be applied and with the same result.  The operation
        $$\cdot\, \,\colon \R \times C \rightarrow C
                                                                                                                \eqno(31)$$
    comes either from \RF(13) or from \RF(17) since multiplying an element of $C$ by a number in either of the two
senses produces the same result.  The zero section is the mapping
        $$O_{\ze} \,\colon Q \rightarrow C \,\colon q \mapsto O_{\zt_P}(O_\zp(q)) = O_{\sT\zp}(O_{\zt_Q}(q)).
                                                                                                                \eqno(32)$$

    There is the mapping
        $$\zq_\zp \,\colon P \fpr{(\zp,\zp)} P \rightarrow \sT P \,\colon (p_0,p) \mapsto \st\,\zz(0)
                                                                                                                \eqno(33)$$
    with
        $$\zz \,\colon \R \rightarrow P \,\colon s \mapsto p_0 + sp.
                                                                                                                \eqno(34)$$
    The image of $\zq_\zp$ is the subbundle
        $$\sV P = \{w \in \sT P;\; \sT\zp(w) = 0\}
                                                                                                                \eqno(35)$$
     of vertical vectors.

    An important example of a double vector fibration is obtained by using the tangent fibration \RF(1) as the vector
fibration \RF(7).  The diagram
    \vskip1mm
        $$\vcenter{\xymatrix@R+3mm @C+2mm{&{\sT\sT Q} \ar[dl]_*{\zt_{\sT Q}} \ar[dr]^*{\sT\zt_Q} & \cr
            \sT Q \ar[dr]^*{\zt_Q} && \sT Q \ar[dl]_*{\zt_Q} \cr
                    & Q &}}
                                                                                                                \eqno(36)$$
    \vskip2mm
    \noindent represents the double fibration.  We follow the costruction of this double fibration described in [9].

    Each element of $\sT\sT Q$ is represented as an equivalence class of mappings
        $$\zq \,\colon \R^2 \rightarrow Q \,\colon (s,t) \mapsto \zq(s,t).
                                                                                                                \eqno(37)$$
    Mappings $\zq$ and $\zq'$ are equivalent if
        $$\rD^{(0,0)}(f \circ \zq')(0,0) = \rD^{(0,0)}(f \circ \zq)(0,0),
                                                                                                                \eqno(38)$$
        $$\rD^{(1,0)}(f \circ \zq')(0,0) = \rD^{(1,0)}(f \circ \zq)(0,0),
                                                                                                                \eqno(39)$$
        $$\rD^{(0,1)}(f \circ \zq')(0,0) = \rD^{(0,1)}(f \circ \zq)(0,0),
                                                                                                                \eqno(40)$$
    and
        $$\rD^{(1,1)}(f \circ \zq')(0,0) = \rD^{(1,1)}(f \circ \zq)(0,0),
                                                                                                                \eqno(41)$$
    for each function $f$ on $Q$.  The symbol $\rD^{(i,j)}$ is used to denote partial derivatives of functions on $\R^2$.
A representative $\zq$ of an element $w \in \sT\sT Q$ is easily constructed by using local coordinates of $w$.  The
equivalence class of $\zq$ will be denoted by $\st^{(1,1)}\zq(0,0)$.  In terms of this representation the mappings
$\zt_{\sT Q}$ and $\sT\zt_Q$, and the composition $\zt_{\sT Q} \circ \sT\zt_Q$ are expressed by
        $$\zt_{\sT Q} \,\colon \sT\sT Q \rightarrow \sT Q \,\colon \st^{(1,1)}\zq(0,0) \mapsto \st\,\zq(0,\,\cdot\,)(0),
                                                                                                                \eqno(42)$$
        $$\sT\zt_Q \,\colon \sT\sT Q \rightarrow \sT Q \,\colon \st^{(1,1)}\zq(0,0) \mapsto \st\,\zq(\,\cdot\,,0)(0),
                                                                                                                \eqno(43)$$
    and
        $$\zt_{\sT Q} \circ \sT\zt_Q \,\colon \sT\sT Q \rightarrow Q \,\colon \st^{(1,1)}\zq(0,0) \mapsto \zq(0,0).
                                                                                                                \eqno(44)$$

    An involution
        $$\zk_Q \,\colon \sT\sT Q \rightarrow \sT\sT Q \,\colon \st^{(1,1)}\zq(0,0) \mapsto \st^{(1,1)}\widetilde\zq(0,0)
                                                                                                                \eqno(45)$$
    is constructed in terms of the above convenient representation of elements of $\sT\sT Q$.  The symbol $\widetilde\zq$
denotes the mapping
        $$\widetilde\zq \,\colon \R^2 \rightarrow Q \,\colon (s,t) \mapsto \zq(t,s).
                                                                                                                \eqno(46)$$
    The diagram
    \vskip2mm
        $$\vcenter{
        \begindc{\commdiag}[1]
        \obj(40,90)[A]{$\sT\sT Q$}
        \obj(0,55)[B]{$\sT Q$}
        \obj(80,35)[C]{$\sT Q$}
        \obj(40,0)[D]{$Q$}
        \obj(200,90)[E]{$\sT\sT Q$}
        \obj(160,55)[F]{$\sT Q$}
        \obj(240,35)[G]{$\sT Q$}
        \obj(200,0)[H]{$Q$}
        \obj(120,0){$\HEQ{50mm}$}
        \obj(80,55){$\HEQ{48mm}$}
        \obj(160,35){$\HEQ{48mm}$}
        \mor{A}{E}[15,15]{$\zk_Q$}
        \mor{A}{B}[10,14]{$\zt_{\sT Q}$}[-1,0]
        \mor{A}{C}[10,9]{$\sT\zt_Q$}[1,0]
        \mor{B}{D}[10,8]{$\zt_Q$}[-1,0]
        \mor{C}{D}[10,8]{$\zt_Q$}[1,0]
        \mor{E}{G}[10,9]{$\zt_{\sT Q}$}[1,0]
        \mor{E}{F}[10,10]{$\sT\zt_Q$}[-1,0]
        \mor{F}{H}[10,8]{$\zt_Q$}[-1,0]
        \mor{G}{H}[10,8]{$\zt_Q$}[1,0]
        \enddc}\;\;\;
                                                                                                                \eqno(47)$$
    \vskip2mm
    \noindent is an isomorphism of double vector fibrations.

        \sect{The tangent functor applied to a dual pair of vector fibrations.}
    Let
    \vskip1mm
        $$\vcenter{
        \begindc{0}[1]
        \obj(000,45)[01]{$E$}   \obj(120,45)[11]{$F$}
        \obj(000,00)[00]{$M$}   \obj(120,00)[10]{$M$}
        \mor{01}{00}[10,8]{$\ze$}[-1,0]
        \mor{11}{10}[10,8]{$\zf$}[-1,0]
        \enddc}
                                                                                                                \eqno(48)$$
    \vskip2mm
    \noindent represent a vector fibration $\ze$ and a dual vector fibration $\zf$. Let
        $$\langle \,\;,\;\rangle \,\colon F \fpr{(\zf,\ze)} E \rightarrow \R
                                                                                                                \eqno(49)$$
    be the pairing.  We have double vector bundle structures in $\sT E$ and $\sT F$.  The tangent fibrations
    \vskip1mm
        $$\vcenter{
        \begindc{0}[1]
        \obj(000,52)[01]{$\sT E$}   \obj(120,52)[11]{$\sT F$}
        \obj(000,00)[00]{$\sT M$}   \obj(120,00)[10]{$\sT M$}
        \mor{01}{00}[10,8]{$\sT\ze$}[-1,0]
        \mor{11}{10}[10,8]{$\sT\zf$}[-1,0]
        \enddc}
                                                                                                                \eqno(50)$$
    \vskip2mm
    \noindent are a dual pair of vector fibrations.  The {\it tangent pairing}
        $$\langle \,\;,\;\rangle^{\ssT} \,\colon \sT F \fpr{(\sT\zf,\sT\ze)} \sT E \rightarrow \R
                                                                                                                \eqno(51)$$
    is constructed from the tangent mapping
        $$\sT\langle \,\;,\;\rangle \,\colon \sT(F \fpr{(\zf,\ze)} E) \rightarrow \sT\,\R
                                                                                                                \eqno(52)$$
    by identifying the space $\sT(F \fpr{(\zf,\ze)} E)$ with $\sT F \fpr{(\sT\zf,\sT\ze)} \sT E$ and composing the tangent
mapping with the second projection of $\sT\,\R = \R^2$ onto $\R$. If $(w,v)$ is an element of $\sT F \fpr{(\sT\zf,\sT\ze)}
\sT E$ and $(\zs,\zr)$ is a curve in $F \fpr{(\zf,\ze)} E$ such that $(w,v) = (\st\,\zs(0),\st\,\zr(0))$, then
        $$\langle w, v\rangle^{\ssT} = \langle \st\,\zs(0), \st\,\zr(0)\rangle^{\ssT} = \rD\langle \zs, \zr\rangle(0).
                                                                                                                \eqno(53)$$
    If $(\zs,\zr)$ is a curve in $F \fpr{(\zf,\ze)} E$, then
        $$\langle \st\,\zs, \st\,\zr\rangle^{\ssT} = \rD\langle \zs, \zr\rangle.
                                                                                                                \eqno(54)$$

    Linearity of the tangent mapping \RF(49) implies linearity of the tangent pairing.  If $(w_1,v_1)$ and $(w_2,v_2)$
are elements of $\sT F \fpr{(\sT\zf,\sT\ze)} \sT E$ such that $\zt_F(w_1) = \zt_F(w_2)$ and $\zt_E(v_1) = \zt_E(v_2)$, then
        $$\langle w_1 + w_2, v_1 + v_2 \rangle^{\ssT} = \langle w_1, v_1 \rangle^{\ssT} + \langle w_2, v_2 \rangle^{\ssT}.
                                                                                                                \eqno(55)$$
    If $(w,v) \in \sT F \fpr{(\sT\zf,\sT\ze)} \sT E$ and $k \in \R$, then
        $$\langle kw, kv\rangle^{\ssT} = k\langle w, v\rangle^{\ssT}.
                                                                                                                \eqno(56)$$

        \sect{Linear and vertical forms on vector bundles.}
    Let $N$ be a differential manifold.  The tangent bundle is denoted by $\sT N$, the tangent fibration is a mapping
$\zt_N \,\colon \sT N \rightarrow N$.  Let $\PR_N^n \sT N$ denote the $n$-fold fibred product of the tangent bundle $\sT
N$ for $n > 0$.  We choose to present a {\it differential n-form} as a differentiable mapping
        $$\zm \,\colon \PR_N^n \sT N \rightarrow \R.
                                                                                                                \eqno(57)$$
    Restricted to a fibre $\PR^n \sT_b N$ an $n$-form is $n$-linear and totally antisymmetric.  A 0-form is a
differentiable function on $N$.  The space of $n$-forms on $N$ will be denoted by $\zF^n(N)$.

    Let
    \vskip1mm
        $$\vcenter{
        \begindc{0}[1]
        \obj(000,45)[01]{$P$}
        \obj(000,00)[00]{$Q$}
        \mor{01}{00}[8,8]{$\zp$}[-1,0]
        \enddc}
                                                                                                                \eqno(58)$$
    \vskip2mm
    \noindent be a vector fibration.  The mapping
        $$\eqalign{
     \PR^n_{Q,P}\sT\zp &\,\colon \PR^n_P \sT P \rightarrow \PR^n_Q \sT Q \cr
    &\,\colon (w^1,\ldots,w^n) \mapsto (\sT\zp(w^1),\ldots,\sT\zp(w^n))}
                                                                                                                \eqno(59)$$
    is again a vector fibration with operations
        $$\eqalign{
    \tdot^n &\,\colon \R \times (\PR^n_P \sT P) \rightarrow \PR^n_P \sT P \cr
    &\,\colon (k,(w^1,\ldots,w^n)) \mapsto (k \tdot w^1,\ldots,k \tdot w^n)}
                                                                                                                \eqno(60)$$
    and
        $$\eqalign{
    \tplus^n &\,\colon (\PR^n_P \sT P) \times_{\PR^n_Q\sT Q} (\PR^n_P \sT P) \rightarrow \PR^n_P \sT P \cr
    &\,\colon ((w_1^1,\ldots,w_1^n),(w_2^1,\ldots,w_2^n)) \mapsto (w_1^1 \tplus w_2^1,\ldots,w_1^n \tplus w_2^n),}
                                                                                                                \eqno(61)$$
    and the zero section
        $$O_{\PR^n_{Q,P}\sT\zp} \,\colon \PR^n_Q \sT Q \rightarrow \PR^n_P \sT P \,\colon (v^1,\ldots,v^n) \mapsto
(O_{\sT\zp}(v^1),\ldots,O_{\sT\zp}(v^n))
                                                                                                                \eqno(62)$$

    An $n$-form
        $$\zm \,\colon \PR^n_P \sT P \rightarrow \R
                                                                                                                \eqno(63)$$
     on $P$ is said to be {\it linear} if it is a function linear on fibres of the fibration $\PR^n_{Q,P}\sT\zp$.  Linear
forms on vector fibrations are examples of more general polynomial forms studied in [10].  Properties of linear forms
follow from properties of polynomial forms established in [10].

    If
    \vskip1mm
        $$\vcenter{\xymatrix@R+2mm @C+8mm{{P} \ar[d]_*{\zp} \ar[r]^*{\zf} &
            P' \ar[d]_*{\zp'} \cr
            Q \ar@{=}[r] & Q}}
                                                                                                                \eqno(64)$$
    \vskip2mm
    \noindent is a vector fibration morphism and $\zm$ is a linear form on $P'$, then $\zf^\*\zm$ is a linear form on $P$.

    The exterior differential of a linear form on $P$ is a linear form on $P$.  A closed linear form is exact.  It is the
exterior differential of a linear form.

    A 1-form $\zm$ on $P$ is said to be {\it vertical} if $\zm(w) = 0$ for each vector $w \in \sT P$ such
that $\sT\zp(w) = 0$.  The pullback $\zf^\*\zm$ of a vertical form is vertical.

        \sect{The Liouville form for a cotangent fibration.}
    The {\it Liouville form} for the cotangent fibration of a manifold $Q$ is the 1-form
        $$\zy_Q \,\colon \sT\sT^\*Q \rightarrow \R
                                                                                                                \eqno(65)$$
    defined by
        $$\zy_Q(w) = \langle \zt_{\sT^\*Q}(w), \sT\zp_Q(w) \rangle\VPD{4pt}_Q.
                                                                                                                \eqno(66)$$

    A local chart
        $$(q^\zk) \,\colon U \rightarrow \R^m
                                                                                                                \eqno(67)$$
    in $U \subset Q$ induces a chart
        $$(q^\zk,p_\zl) \,\colon U^\* \rightarrow \R^{2m}
                                                                                                                \eqno(68)$$
    in $U^\* = \zp_Q^{-1}(U)$ such that
        $$\zy_Q|U^\* = p_\zk \rd q^\zk.
                                                                                                                \eqno(69)$$
    From the local expression
        $$\zw_Q|U^\* = \rd p_\zk \wedge \rd q^\zk
                                                                                                                \eqno(70)$$
    we deduce that the 2-form $\zw_Q = \rd \zy_Q$ is non degenerate.  It follows that the cotangent bundle $\sT^\*Q$ with
the 2-form $\zw_Q = \rd\zy_Q$ form a symplectic manifold $(\sT^\*Q,\zw_Q)$.

    The Liouville form is vertical since $\sT\zp_Q(w) = 0$ implies
        $$\zy_Q(w) = \langle \zt_{\sT^\*Q}(w), \sT\zp_Q(w) \rangle\VPD{4pt}_Q = 0.
                                                                                                                \eqno(71)$$

    If $w_1 \in \sT P$ and $w_2 \in \sT P$ are vectors such that $\sT\zp_Q(w_2) = \sT\zp_Q(w_1) = v$, then
        $$\eqalign{
    \zy_Q(w_1 \tplus w_2)  =& \langle \zt_{\sT^\*Q}(w_1 \tplus w_2), \sT\zp_Q(w_1 \tplus w_2) \rangle\VPD{4pt}_Q \cr
            &= \langle \zt_{\sT^\*Q}(w_1) + \zt_{\sT^\*Q}(w_2), v \rangle\VPD{4pt}_Q \cr
            &= \langle \zt_{\sT^\*Q}(w_1), \sT\zp_Q(w_1) \rangle\VPD{4pt}_Q + \langle \zt_{\sT^\*Q}(w_2), \sT\zp_Q(w_2)
\rangle\VPD{4pt}_Q \cr
            &= \zy_Q(w_1) + \zy_Q(w_2).}
                                                                                                                \eqno(72)$$
    For each vector $w \in \sT P$ and each number $k$ we have
        $$\eqalign{
    \zy_Q(k \tdot w)  =& \langle \zt_{\sT^\*Q}(k \tdot w), \sT\zp_Q(k \tdot w) \rangle\VPD{4pt}_Q \cr
            &= \langle k\zt_{\sT^\*Q}(w), \sT\zp_Q(w) \rangle\VPD{4pt}_Q \cr
            &= k \langle \zt_{\sT^\*Q}(w), \sT\zp_Q(w) \rangle\VPD{4pt}_Q \cr
            &= k\zy_Q(w).}
                                                                                                                \eqno(73)$$
    It follows that the Liouville form is linear.

        \sect{Liouville structures.}
    A {\it Liouville structure} is a vector fibration isomorphism
    \vskip1mm
        $$\vcenter{\xymatrix@R+2mm @C+8mm{{P} \ar[d]_*{\zp} \ar[r]^*{\za} &
            \sT^\*Q \ar[d]_*{\zp_Q} \cr
            Q \ar@{=}[r] & Q}}
                                                                                                                \eqno(74)$$
    \vskip2mm
    \noindent This is a preliminary definition.  Alternative definitions will be formulated.  Liouville structures form a
category.  We will define morphisms and functors in this category after we have introduced the alternative definitions.

    The following structures in the vector fibration
    \vskip1mm
        $$\vcenter{
        \begindc{0}[1]
        \obj(000,45)[01]{$P$}
        \obj(000,00)[00]{$Q$}
        \mor{01}{00}[8,8]{$\zp$}[-1,0]
        \enddc}
                                                                                                                \eqno(75)$$
    \vskip2mm
    \noindent are derived from the Liouville structure \RF(74).

        \List\vskip1mm
    \item {1)} A bilinear non degenerate pairing
        $$\langle \,\; ,\;\rangle \,\colon P \fpr{(\zp,\zt_Q)} \sT Q \rightarrow \R \,\colon (p,v) \mapsto \langle p,
v\rangle = \langle \za(p), v\rangle\VPD{4pt}_Q
                                                                                                                \eqno(76)$$
    is established.
    \item {2)} A non degenerate linear 2-form $\zw = \za^\*\zw_Q$ is introduced.  The structure is represented by the
diagram
    \vskip1mm
        $$\vcenter{
        \begindc{0}[1]
        \obj(000,45)[01]{$(P,\zw)$}
        \obj(000,00)[00]{$Q$}
        \mor{01}{00}[8,8]{$\zp$}[-1,0]
        \enddc}
                                                                                                                \eqno(77)$$
    \vskip2mm
    \item {3)} A 1-form $\zy$ on $P$ is defined as the pullback $\za^\*\zy_Q$ of the canonical Liouville form $\zy_Q$. The
form $\zy$ is vertical and linear.  The differential $\rd\zy = \za^\*\zw_Q$ is non degenerate.  The structure is
represented by the diagram
    \vskip1mm
        $$\vcenter{
        \begindc{0}[1]
        \obj(000,45)[01]{$(P,\zy)$}
        \obj(000,00)[00]{$Q$}
        \mor{01}{00}[8,8]{$\zp$}[-1,0]
        \enddc}
                                                                                                                \eqno(78)$$
    \vskip2mm
        \endList

\vskip2mm
    We show that each of these structures separately defines a Liouville structure for the vector fibration \RF(75).
\vskip2mm
        \ssect{Liouville structures constructed from duality.}
    Let the vector fibration \RF(75) form a dual pair with the tangent fibration
    \vskip1mm
        $$\vcenter{
        \begindc{0}[1]
        \obj(000,45)[01]{$\sT Q$}
        \obj(000,00)[00]{$Q$}
        \mor{01}{00}[8,8]{$\zt_Q$}[-1,0]
        \enddc}
                                                                                                                \eqno(79)$$
    \vskip2mm
    \noindent with a bilinear non degenerate pairing
        $$\langle \;\, ,\;\rangle \,\colon P \fpr{(\zp,\zt_Q)} \sT Q \rightarrow \R.
                                                                                                                \eqno(80)$$
    A mapping $\za \,\colon P \rightarrow \sT^\*Q$ is introduced.  It is characterized by
        $$\langle \za(p), v\rangle\VPD{4pt}_Q = \langle p, v\rangle
                                                                                                                \eqno(81)$$
    for $p \in P$ and each $v \in \sT Q$ such that $\zt_Q(v) = \zp(p)$.  The diagram
    \vskip1mm
        $$\vcenter{\xymatrix@R+2mm @C+10mm{{P} \ar[d]_*{\zp} \ar[r]^*{\za} &
            \sT^\*Q \ar[d]_*{\zp_Q} \cr
            Q \ar@{=}[r] & Q}}
                                                                                                                \eqno(82)$$
    \vskip2mm
    \noindent is vector fibration morphism.  Let $q$ be a point in $Q$ and let $\za_q \,\colon P_q \rightarrow \sT_q^\*Q$
be the mapping induced by the restriction of $\za$ to the fibre $P_q = \zp^{-1}(q)$.  If $\za_q(p) = O_{\zp_Q}(q)$, then
$\langle p, v\rangle = \langle \za_q(p), v\rangle\VPD{4pt}_Q = 0$ for each $v \in \sT_q Q$.  It follows that $p = O_\zp(q)$
since the pairing \RF(80) is non degenerate.  We have shown that $\za_q$ is injective.  If ${\rm im}(\za_q) \neq \sT_q^\*Q$,
then there is a vector $v \neq 0$ in $\sT_q Q$ such that $\langle p, v\rangle = \langle \za(p), v\rangle\VPD{4pt}_Q = 0$
for each $p \in P_q$.  This is not possible since the pairing \RF(80) is non degenerate.  We have shown that $\za_q$ is
surjective. The diagram \RF(82) is a Liouville structure.

        \ssect{Liouville structures derived from linear symplectic structures.}
    Let $\zw$ be a symplectic form on a manifold $P$.  If $P$ is the total space of a vector fibration \RF(75) and the
form $\zw$ is linear, then there is a unique Liouville structure \RF(74) such that $\zw = \za^\*\zw_Q$.  This Liouville
structure is said to be {\it a Liouville structure} for the symplectic manifold $(P,\zw)$.  Several Liouville structures for
the same symplectic manifold may be considered.

    Let
    \vskip1mm
        $$\vcenter{
        \begindc{0}[1]
        \obj(000,45)[01]{$(P,\zw)$}
        \obj(000,00)[00]{$Q$}
        \mor{01}{00}[8,8]{$\zp$}[-1,0]
        \enddc}
                                                                                                                \eqno(83)$$
    \vskip2mm
    \noindent be a Liouville structure specified in terms of a linear symplectic form.  A pairing is defined by
        $$\langle \,\;, \;\rangle \,\colon P \fpr{(\zp,\zt_Q)} \sT Q \rightarrow \R \,\colon (p,v) \mapsto
\zw(\zq_\zp(O_\zp(\zp(p)),p),\sT O_{\zt_Q}(v)).
                                                                                                                \eqno(84)$$
    We show that the pairing is non degenerate.  At each $q \in Q$ we consider the restriction of $\zw$ to the space
        $$\sT_{O_\zp(q)}P \times \sT_{O_\zp(q)}P.
                                                                                                                \eqno(85)$$
    This restriction will be denoted by $\zw_q$.  The space $\sT_{O_\zp(q)}P$ is the sum $V_q + H_q$ of the subspace $V_q$
of vertical vectors and the space
        $$H_q = \sT O_\zp(\sT_q Q).
                                                                                                                \eqno(86)$$
    Of the four components
        $$\zw_q^{v,v} = \zw|(V_q \times V_q),
                                                                                                                \eqno(87)$$
        $$\zw_q^{v,h} = \zw|(V_q \times H_q),
                                                                                                                \eqno(88)$$
        $$\zw_q^{v,h} = \zw|(H_q \times V_q),
                                                                                                                \eqno(89)$$
    and
        $$\zw_q^{h,h} = \zw|(H_q \times H_q)
                                                                                                                \eqno(90)$$
    the second appears in the definition \RF(84) of the pairing since $\zq_\zp\left(O_\zp(\zp(p),p)\right) \in V_q$ and
$\sT O_{\zt_Q}(v) \in H_q$ if $\zp(p) = q$.  Refering to the diagram \RF(28) we observe that $V_q = C_q = \ze^{-1}(q)$.
Hence, if $w \in V_q$, then $k \,\tdot w = kw$.  If $w_1 \in V_q$ and $w_2 \in V_q$, then
        $$\zw(kw_1,kw_2) =  k^2\zw(w_1,w_2)
                                                                                                                \eqno(91)$$
    and
        $$\zw(kw_1,kw_2) = \zw(k \tdot w_1,k \tdot w_2) = k\zw(w_1,w_2)
                                                                                                                \eqno(92)$$
    for each $k \in \R$.  It follows that $\zw(kw_1,kw_2) = 0$.  Hence, $\zw_q^{v,v} = 0$.  We observe that
        $$H_q \times H_q = O_{\times^2_{Q,P}\sT\zp}(\sT_q Q \times \sT_q Q).
                                                                                                                \eqno(93)$$
    It follows immediately that $\zw_q^{h,h} = 0$ since $\zw$ is
linear.  The component $\zw_q^{v,h}$ must be non degenerate since $\zw_q$ is non degenerate.  We have shown that the
pairing \RF(84) is non degenerate.  A Liouville structure
    \vskip1mm
        $$\vcenter{\xymatrix@R+2mm @C+10mm{{P} \ar[d]_*{\zp} \ar[r]^*{\za} &
            \sT^\*Q \ar[d]_*{\zp_Q} \cr
            Q \ar@{=}[r] & Q}}
                                                                                                                \eqno(94)$$
    \vskip2mm
    \noindent is constructed from the pairing \RF(84).  The mapping $\za$ is characterized by
        $$\langle \za(p), v\rangle\VPD{4pt}_Q = \langle p, v\rangle
                                                                                                                \eqno(95)$$
    for each $p \in P$ and each $v \in \sT Q$ such that $\zt_Q(v) = \zp(p)$.  It follows from results established in
[10] that
        $$\za^{\*}\zw_Q = \zw.
                                                                                                                \eqno(96)$$

        \ssect{Liouville structures derived from Liouville forms.}
    Let
    \vskip1mm
        $$\vcenter{
        \begindc{0}[1]
        \obj(000,45)[01]{$(P,\zy)$}
        \obj(000,00)[00]{$Q$}
        \mor{01}{00}[8,8]{$\zp$}[-1,0]
        \enddc}
                                                                                                                \eqno(97)$$
    \vskip2mm
    \noindent be a Liouville structure specified in terms of a linear vertical 1-form on $\zy$ such that the form $\rd\zy$
is non degenerate.  The form $\zy$ is called a {\it Liouville form} on $P$.  Using the linear symplectic form $\zw =
\rd\zy$ we construct a Liouville structure as in the preceding subsection.  We know already that
        $$\za^{\*}\rd\zy_Q = \rd\zy.
                                                                                                                \eqno(98)$$
    We have to show that
        $$\za^{\*}\zy_Q = \zy.
                                                                                                                \eqno(99)$$
    The form $\zy - \za^{\*}\zy_Q$ is linear, vertical, and closed.  A closed linear form is the differential of a linear
form.  It follows that there is a linear function $f \,\colon P \rightarrow \R$ such that $\zy - \za^{\*}\zy_Q = \rd f$.
The vertical vector $\zq_\zp(O_\zp(\zp(p)),p)$ associated with $p \in P$ is represented by the curve
        $$\zz \,\colon \R \rightarrow P \,\colon s \mapsto sp.
                                                                                                                \eqno(100)$$
    We have
        $$f(p) = {{\rd}\over{\rd s}} f(sp)\big|_{s=0} = \rd f(\zq_\zp(O_\zp(\zp(p)),p)) = 0
                                                                                                                \eqno(101)$$
    since $f$ is linear and $\rd f$ is vertical.  We have obtained the equality \RF(99).

    Starting with a linear, vertical form $\zy$ and its differential $\zw$ we have costructed the pairing \RF(84) and
the Liouville structure \RF(94).  For $p \in P$ and $v \in \sT Q$ such that $\zt_Q(v) = \zp(p)$ we choose a vector $w
\in \sT P$ such that $\zt_P(w) = p$ and $\sT\zp(w) = v$.  From
        $$\eqalign{
            \zy(w) &= \za^{\*}\zy_Q \cr
        &= \zy_Q(\sT\za(w)) \cr
        &= \langle \zt_{\sT^\*Q}(\sT\za(w)), \sT\zp_Q(\sT\za(w))\rangle \cr
        &= \langle \za(p), v\rangle \cr
        &= \langle p, v\rangle}
                                                                                                                \eqno(102)$$
    it follows that the pairing \RF(84) can be defined directly from the form $\zy$.  It is defined by
        $$\langle \,\;, \;\rangle \,\colon P \fpr{(\zp,\zt_Q)} \sT Q \rightarrow \R \,\colon (p,v) \mapsto \zy(w),
                                                                                                                \eqno(103)$$
    with a choice of the vector $w$ described above.  Although this definition is simpler than \RF(84) the derivation of
the required properties of the pairing is more difficult with this definition.

    For reasons of simplicity of notation it is convenient to represent a Liouvile structure by the diagram
    \vskip1mm
        $$\vcenter{
        \begindc{0}[1]
        \obj(000,45)[01]{$(P,\zy)$}
        \obj(000,00)[00]{$Q$}
        \mor{01}{00}[8,8]{$\zp$}[-1,0]
        \enddc}
                                                                                                                \eqno(104)$$
    \vskip2mm
    \noindent with the Liouville form $\zy$ rather than the diagram \RF(1) with the isomorphism $\za$.

\def\rl#1{\;#1\;}
 We should mention here that Liouville forms, but in a different context were studied by Libermann in [4].
        \sect{Relations.}
        A {\it relation} from a set $A$ to a set $B$ is a triple $(B,A,R)$, where $R$ is a subset of the product $B\times
A$.  The set $R$ is called the {\it graph} of a relation $\zr = (B,A,R)$.  We write $\zr\,\colon A \rightarrow B$ to
indicate that $\zr$ is a relation from $A$ to $B$.  We write $b\rl\zr a$ to indicate that the pair $(b,a)\in B\times A$ is
an element of the graph $R$.  In this case we also say that the elements $a$ and $b$ are in the relation $\zr$.  The above
defined concept of a relation conforms to the concept of a morphism in category theory.

    The {\it composition} of a relation $\zr\,\colon A \rightarrow B$ with a relation $\zs\,\colon B \rightarrow C$ is a
relation $\zs\circ\zr\,\colon A \rightarrow C$ defined by
        $$c\, (\zs\circ\zr)\,a \;\;{\rm if\; there\; is\; an\; element}\;\; b\in B \;\;{\rm such\; that}\;\; c\;\zs\, b
\;\;{\rm and}\;\; b\;\zr\; a.
                                                                                                                \eqno(105)$$
    The {\it transpose} of a relation $\zr = (B,A,R)$ is the relation $\zr^t = (A,B,R^t)$ with the graph
        $$R^t = \left\{(a,b) \in A \times B ;\; (b,a) \in R \right\}.
                                                                                                                \eqno(106)$$

    A relation $\za \,\colon A \rightarrow B$ is called a {\it mapping} if for each element $a\in A$ there is a unique
element $b\in B$ such that $b\rl\za a$.  This unique element $b$ is denoted by $\za(a)$.  If $\za$ is a mapping, then $b =
\za(a)$ has the same meaning as $b\rl\za a$.  If $\zb\,\colon B \rightarrow C$ and $\za\,\colon A \rightarrow B$ are
mappings, then the composition $\zb\circ\za$ is a mapping from $A$ to $C$ defined by
        $$(\zb\circ\za)(a) = \zb(\za(a)).
                                                                                                                \eqno(107)$$
    The transpose of a mapping is not usually a mapping.  The transpose of a bijective mapping $\za$ is its inverse
$\za^{-1}$.

        A {\it relation} $\zr = (N,M,R)$ from a differential manifold $M$ to a differential manifold $N$ is said to be {\it
differentiable} if the graph $R$ is a submanifold of $N\times M$.  The transpose of a differentiable relation is
differentiable.  The composition of differentiable relations is not necessarily differentiable.  Differential manifolds and
differentiable relations do not form a category.

    Let $C$ be a submanifold of a differential manifold $M$ and let $\zk \,\colon C \rightarrow N$ be a differential
fibration.  Let $\zi_C \,\colon C \rightarrow M$ be the canonical injection.  The composition $\zr = \zk \circ \zi_C{}^t$ is
a differentiable relation from $M$ to $N$.  The graph of $\zr$ is the submanifold
        $$R = \left\{(x,y) \in M \times N ;\;x \in C,  y = \zk(x)\right\}.
                                                                                                                \eqno(108)$$
    The relation $\zr$ is called a {\it differentiable reduction}.  Let $\zs = \zl \circ \zi_D{}^t \,\colon N \rightarrow
T$ be a differentiable reduction constructed from a submanifold $D \subset N$ and a differential fibration $\zl \,\colon D
\rightarrow T$.  The set $E = \zi_C(\zk^{-1}(D))$ is a submanifold of $M$.  The mapping
        $$\zm \,\colon E \rightarrow T \,\colon x \mapsto \zl(\zk(x))
                                                                                                                \eqno(109)$$
    is a differential fibration.  The composition $\zs \circ \zr$ is the differentiable reduction $\zm \circ \zi_E{}^{-1}$.
Differential manifolds and differentiable reductions form a category.  The category of differentiable mappings is a
subcategory of the category of differentiable reductions.

    A {\it differential fibration morphism} is a commutative diagram
    \vskip1mm
        $$\vcenter{\xymatrix@R+3mm @C+10mm{{P} \ar[d]_*{\zp} \ar[r]^*{\zr} &
            P' \ar[d]_*{\zp'} \cr
            Q \ar[r]^*{\zs} & Q'}}
                                                                                                                \eqno(110)$$
    \vskip2mm
    \noindent where
    \vskip1mm
        $$\vcenter{
        \begindc{0}[1]
        \obj(000,45)[01]{$P$}   \obj(120,45)[11]{$P'$}
        \obj(000,00)[00]{$Q$}   \obj(120,00)[10]{$Q'$}
        \mor{01}{00}[10,8]{$\zp$}[-1,0]
        \mor{11}{10}[10,8]{$\zp'$}[-1,0]
        \enddc}
                                                                                                                \eqno(111)$$
    \vskip2mm
    \noindent are differential fibrations and $\zr$ and $\zs$ are differentiable relations.

    A {\it vector fibration morphism} is a differential fibration morphism \RF(110).  The fibrations \RF(111) are
vector fibrations and for each $(q',q) \in \gr(\zs)$ the set $\gr(\zr) \cap (P'_{q'} \times P_{q})$ is a vector subspace of
$P'_{q'} \times   P_{q}$.

    Let $(P,\zw)$ be a symplectic manifold and let $\zb \,\colon \sT P \rightarrow \sT^\*P$ be the natural isomorphism
characterized by
        $$\langle w, \zb(v) \rangle = \langle v \wedge w, \zw \rangle
                                                                                                                \eqno(112)$$
    for each $v \in \sT P$ and each $w \in \sT P$ such that $\zt_P(w) = \zt_P(v)$. Let $V \subset \sT_pP$ be a vector
subspace.  The {\it polar} $V^\polar$ is the subspace
        $$\left\{q \in \sT^\*_pP ;\;  \all{v \in V} \langle v, q \rangle = 0 \right\}.
                                                                                                                \eqno(113)$$
    We denote by $V^{\srP}$ the {\it symplectic polar}
        $$\zb^{-1}(V^\polar) = \left\{w \in \sT_p P ;\; \all{v \in V} \langle v \wedge w,\zw \rangle = 0 \right\}.
                                                                                                                \eqno(114)$$
    If $C \subset P$ is a submanifold, then $\sT^{\srP} C$ will denote the set
        $$\bigcup_{p \in C} \left(\sT_p C\right)^{\srP}.
                                                                                                                \eqno(115)$$
    We recall that a submanifold $C \subset P$ is said to be {\it isotropic} if $\sT^{\srP} C \supset \sT C$.  A
submanifold $C \subset P$ is said to be {\it coisotropic} if $\sT^{\srP} C \subset \sT C$.  The set $\sT^{\srP} C$ is
called the {\it characteristic distribution} of a coisotropic submanifold $C \subset P$.  The characteristic distribution
is Frobenius integrable.  Its integral foliation is called the {\it characteristic foliation} of $C$.  The maximum
dimension of an isotropic submanifold is a half of the dimension of $P$.  A {\it Lagrangian submanifold} of a symplectic
manifold $(P,\zw)$ is an isotropic submanifold $C \subset P$ of maximum dimension.  The equality $\sT^{\srP} C = \sT C$ is
an equivalent characterization of a Lagrangian submanifold.

    A {\it symplectic relation} from a symplectic manifold $(P,\zw)$ to a symplectic manifold $(P',\zw')$ is a
differentiable relation $\zr$ from $P$ to $P'$ with the additional property that the graph of $\zr$ is a Lagrangian
submanifold of the symplectic manifold $(P' \times P,\zw' \ominus \zw)$, where the form $\zw' \ominus \zw$ is defined by
        $$\zw' \ominus \zw = pr'{}^\* \zw' - pr^\* \zw,
                                                                                                                \eqno(116)$$
    where $pr \,\colon P' \times P \rightarrow P$ and $pr' \,\colon P' \times P \rightarrow P'$ are the canonical
projections. The transpose of a symplectic relation is a symplectic relation.  The composition of symplectic relations is
not necessarily differentiable and if it is diferentiable is not necessarily symplectic.

    Let $(P,\zw)$ and $(P',\zw')$$(P,\zw)$ be symplectic manifolds and let $\zr = \zk \circ \zi_C{}^{-1}$ be a
differentiable reduction constructed from a submanifold $C \subset P$ and a differential fibration $\zk \,\colon C
\rightarrow P'$.  If $C$ is a coisotropic submanifold of $(P,\zw)$, fibres of $\zk$ are characteristics of $C$, and
$\zk^{\*}\zw' = \zi_C{}^{\*}\zw$ then $\zr$ is a symplectic relation.  This relation is called a {\it symplectic
reduction}.  The composition of symplectic reductions is a symplectic reduction.  The only symplectic relations which are
actually mappings are diffeomorphisms known as {\it symplectomorphisms}.

    Symplectic manifolds and symplectic reductions form a category.  Attempts have been made to extend this category to a
wider class of symplectic relations (unpublished results of S. Zakrzewski).  One should probably use symplectic relations without insisting that they form a
category.

        \sect{Liouville structure morphisms.}
        \Proposition{
    Let $\zp \,\colon P \rightarrow Q$ be a vector fibration.  If $K$ is a closed submanifold of $P$ such that for each $q
\in C =\zp(K)$ the intersection $K_q = K \cap P_q$ of $K$ with $P_q = \zp^{-1}(q)$ is a vector subspace of $P_q$, then $C$
is a submanifold of $Q$ and the dimension of $K_q$ is locally constant and the mapping
        $$\overline\zp \,\colon K \rightarrow C \,\colon p \mapsto \zp(p)
                                                                                                                \eqno(117)$$
    is a vector fibration.
        }
        \Proof{
$\ker(\sT\zp|T_p K) = \sT P_q \cap \sT_p K$ is equal to $\sT_p K_q$.  A vector $v \in \ker(\sT\zp|T_p K)$ is in $\sT P_p$.
It has a representation $v = \zq_\zp(p,\zd p)$ in terms of the vector $p$ and a unique vector $\zd p \in P_q$.  The curve
        $$\zz \,\colon \R \rightarrow P \,\colon s \mapsto p + s\zd p
                                                                                                                \eqno(118)$$
     can be used as an integral curve of $v$.  Being in $\sT_p K$ the vector $v$ is the tangent vector $\st\,\zh(0)$ of a
curve
        $$\zh \,\colon \R \rightarrow K
                                                                                                                \eqno(119)$$
    with $\zh(0) = p$.  Both integral curves will be used.  For each $k \neq 0$ we consider the vector $k \tdot v$.  The
curve
        $$k \cdot \zh \,\colon \R \rightarrow K \,\colon s \mapsto k\cdot \zh(s)
                                                                                                                \eqno(120)$$
    with $(k \cdot \zh)(0) = k \cdot \zh(0) = k \cdot p$ is an integral curve of $k \tdot v$.  It follows that $k \tdot v$
is in $\sT_{k \cdot p} K$.  The curve
        $$k \cdot \zz \,\colon \R \rightarrow P \,\colon s \mapsto k \cdot p + sk \cdot \zd p
                                                                                                                \eqno(121)$$
    is again an integral curve of $k \tdot v$.  It follows that $k \tdot v$ is in $\sT P_q$.  Hence $k \tdot v$ is in
$\ker(\sT\zp|\sT_{k \cdot p}K)$.  The vector $k^{-1}(k \tdot v)$ is also in $\ker(\sT\zp|\sT_{k \cdot p}K)$.  The curve
        $$\R \rightarrow P \,\colon s \mapsto k \cdot p + s \cdot \zd p
                                                                                                                \eqno(122)$$
    is an integral curve of this vector.  In the limit of $k \rightarrow 0$ we have
        $$v_0 = \zq_\zp(O_\zp(q),\zd p) \in \ker(\sT\zp|\sT_{O_\zp(q)}K) = \sT P_q \cap \sT_{O_\zp(q)}K.
                                                                                                                \eqno(123)$$
    The vector $v_0$ is tangent to $K$.  It has an integral curve
        $$\zx \,\colon \R \rightarrow K.
                                                                                                                \eqno(124)$$
    for each $k > 0$ we introduce the modified curve
        $$\zx_k \,\colon \R \rightarrow K \,\colon s \mapsto k^{-1}\zx(ks).
                                                                                                                \eqno(125)$$
    For a function $f \,\colon P \rightarrow \R$ constant on fibres of $\zp$ we have
        $$\rD(f \circ \zx)(0) = 0
                                                                                                                \eqno(126)$$
    since $\zx$ represents a vertical vector, and
        $$(f \circ \zx_k)(s) = (f \circ \zx)(ks)
                                                                                                                \eqno(127)$$
    hence,
        $$\rD(f \circ \zx_k)(0) = k\rD(f \circ \zx)(0) = 0.
                                                                                                                \eqno(128)$$
    For a function $g \,\colon P \rightarrow \R$ linear on fibres of $\zp$ we have
        $$(g \circ \zx_k)(s) = k^{-1}(g \circ \zx)(ks)
                                                                                                                \eqno(129)$$
    and
        $$\rD(g \circ \zx_k)(0) = \rD(g \circ \zx)(0).
                                                                                                                \eqno(130)$$
    We have shown that for each $k$ the curve $\zx_k$ is an integral curve of $v_0$.  From
        $$\lim_{k \rightarrow 0} (f \circ \zx_k)(s) = \lim_{k \rightarrow 0}(f \circ \zx)(ks) = (f \circ \zx)(0)
                                                                                                                \eqno(131)$$
    and
        $$\lim_{k \rightarrow 0} (g \circ \zx_k)(s) = \lim_{k \rightarrow 0}k^{-1}(g \circ \zx)(ks) = s\rD(g \circ \zx)(0)
                                                                                                                \eqno(132)$$
    it follows that the family of curves $\zx_k$ converges almost uniformly to an integral curve $\zx_0 \,\colon \R
\rightarrow K$ of $v_0$.  The equality
        $$(f \circ \zx_0)(s) = (f \circ \zx)(0)
                                                                                                                \eqno(133)$$
    for each function $f$ constant on fibres of $\zp$ indicates that the image of $\zx_0$ is in $K_q$.  It follows that
$v_0 = \zq_\zp(O_\zp(q),\zd p)$ is in $\sT_{O_\zp(q)}K_q$, $\zd p$ is in $K_q$, and $v = \zq_\zp(p,\zd p)$ is in
$\sT_pK_q$.  We have shown that
        $$\ker(\sT\zp|T_p K) = \sT_p K_q.
                                                                                                                \eqno(134)$$
\dacapo
    It follows from \RF(134) that
        $$\dim K_q + \dim(\im(\sT\zp|T_p K)) = \dim(\ker(\sT\zp|T_p K)) + \dim(\im(\sT\zp|T_p K)) = \dim(\sT_p K) = \dim K.
                                                                                                                \eqno(135)$$
    The set $C = \zp(K) \subset Q$ is not assumed to be a submanifold.  At each $q \in C$ we have the cone $\sT_q C$ of
vectors in $\sT_q Q$ which have integral curves in $C$.  We observe that
        $$\im(\sT\zp|T_p K) \subset \sT_q C
                                                                                                                \eqno(136)$$
    and, in particular,
        $$\im(\sT\zp|T_{O_\zp(q)} K) \subset \sT_q C.
                                                                                                                \eqno(137)$$
    Let $u$ be a vector in $\sT_q C$.  We have $\sT O_\zp(u) \in \sT_{O_\zp(q)}K$ and $u = \sT\zp(\sT O_\zp(u))$.  It
follows that
        $$\sT_q C \subset \im(\sT\zp|T_{O_\zp(q)} K).
                                                                                                                \eqno(138)$$
    Hence,
        $$\im(\sT\zp|T_{O_\zp(q)} K) = \sT_q C
                                                                                                                \eqno(139)$$
    is a vector space and
        $$\dim\sT_q C = \dim(\im(\sT\zp|T_{O_\zp(q)} K)).
                                                                                                                \eqno(140)$$
    The equality \RF(135) implies the equality
        $$\dim(\im(\sT\zp|T_p K)) = \dim K - \dim K_q = \dim(\im(\sT\zp|T_{O_\zp(q)} K)).
                                                                                                                \eqno(141)$$
    From the inclusion \RF(136), the equality \RF(135), and
        $$\dim\sT_q C = \dim(\im(\sT\zp|T_p K))
                                                                                                                \eqno(142)$$
    we conclude that
        $$\im(\sT\zp|T_p K) = \sT_q C
                                                                                                                \eqno(143)$$
    and
        $$\dim K_q + \dim \sT_q C = \dim K.
                                                                                                                \eqno(144)$$
\dacapo
    Let the vector space $F$ be a typical fibre of the fibration $\zp$ and let
        $$\zk \,\colon \zp^{-1}(U) \rightarrow F
                                                                                                                \eqno(145)$$
    ba a local trivialization of $\zp$ defined in an open neighbourhood $U$ of $q$.  Inclusions
        $$\sT O_\zp(\sT_q C) \subset \sT_{O_\zp(q)}K
                                                                                                                \eqno(146)$$
    and
        $$\sT O_\zp(\sT_q C) \subset \ker(\sT_{O_\zp(q)}\zk),
                                                                                                                \eqno(147)$$
    and the equality
        $$\ker(\sT_{O_\zp(q)}\zk) \cap \sT_{O_\zp(q)}K = \ker(\sT_{O_\zp(q)}\zk|\sT_{O_\zp(q)}K)
                                                                                                                \eqno(148)$$
    result in the inclusion
        $$\sT O_\zp(\sT_q C) \subset \ker(\sT_{O_\zp(q)}\zk|\sT_{O_\zp(q)}K).
                                                                                                                \eqno(149)$$
    The inequality
        $$\dim(\sT_q C) = \dim(\sT O_\zp(\sT_q C)) \leqs \dim(\ker(\sT_{O_\zp(q)}\zk|\sT_{O_\zp(q)}K))
                                                                                                                \eqno(150)$$
    follows.  The inequality
        $$\dim(\im(\sT_{O_\zp(q)}\zk|\sT_{O_\zp(q)}K)) \geqs \dim(\sT_{O_\zp(q)}\zk(\sT_{O_\zp(q)}K_q)) =
\dim(\sT_{O_\zp(q)}K_q) = \dim K_q
                                                                                                                \eqno(151)$$
    follows
        $$\sT_{O_\zp(q)}\zk(\sT_{O_\zp(q)}K_q) \subset \sT_{O_\zp(q)}\zk(\sT_{O_\zp(q)}K).
                                                                                                                \eqno(152)$$
    We have
        $$\dim K_q + \dim \sT_q C = \dim(\im(\sT_{O_\zp(q)}\zk|\sT_{O_\zp(q)}K)) +
\dim(\ker(\sT_{O_\zp(q)}\zk|\sT_{O_\zp(q)}K))
                                                                                                                \eqno(153)$$
    since
        $$\dim(\im(\sT_{O_\zp(q)}\zk|\sT_{O_\zp(q)}K)) + \dim(\ker(\sT_{O_\zp(q)}\zk|\sT_{O_\zp(q)}K)) = \dim K
                                                                                                                \eqno(154)$$
    and
        $$\dim K_q + \dim \sT_q C = \dim K.
                                                                                                                \eqno(155)$$
    The equality
        $$\dim \sT_q C = \dim(\ker(\sT_{O_\zp(q)}\zk|\sT_{O_\zp(q)}K))
                                                                                                                \eqno(156)$$
    is derived from \RF(153), \RF(150), and \RF(151).
\dacapo
    The function
        $$C \rightarrow \N \,\colon q \mapsto \dim \sT_q C
                                                                                                                \eqno(157)$$
    is semi-continuous from above due to \RF(156).  It is also semi-continuous from below due to \RF(140).  It is
continuous on the connected set $C$ and, therefore, constant.  A consequence of this result together with \RF(142) is
that the Constant Rank Theorem applies to the mapping $\zp|K$.  It follows that $C \subset Q$ is a submanifold.  The
dimension of $K_q$ is the same at each $q \in C$ due to \RF(155).  We construct a local trivialization of the
projection $\overline\zp$ starting with a trivialization \RF(145) in a neighbourhood $U$ of a point $q \in C$.  The
typical fibre will be the vector space $\overline F = \zk(K_q)$.  We choose a linear epimorphism $\zq \,\colon F \rightarrow
\overline F$ and introduce the trivialization $\overline\zk \,\colon \zp^{-1}(U \cap C) \rightarrow \overline F \,\colon p
\mapsto \zq(\zk(p))$.  The neighbourhood $U$ is assumed small enough so that the mapping
        $$\zp^{-1}(U \cap C) \rightarrow (U \cap C) \times \overline F \,\colon p \mapsto (\zp(p),\overline\zk(p))
                                                                                                                \eqno(158)$$
    is a diffeomorphism.
        }
This proposition is very close to recent results of Grabowski and Rotkiewicz in [2], concerning vector bundle structures on a manifold. 
    Let $C$ be a submanifold of a manifold $Q$.  The tangent set $\sT C$ is a subbundle of the restriction $\sT_C Q$ of the
tangent bundle $\sT Q$ to $C$.  The {\it anihilator} or the {\it polar} of $\sT_C Q$ is the subbundle
        $$\sT^\polar C = \left\{p \in \sT^\*Q ;\; q = \zp_Q(p) \in C, \all{v \in \sT_q C} \langle p, v\rangle =0 \right\}
                                                                                                                \eqno(159)$$
    of $\sT^\*_C Q$.

        \Proposition{
    If $K$ is a submanifold of the cotangent bundle $\sT^\*Q$ with the properties
        \List
    \item {\rm 1)} the dimension of $K$ is equal to the dimension of $Q$,
    \item {\rm 2)} for each $q \in C = \zp_Q(K)$ the intersection $K_q = K \cap \sT^\*_q Q$ of $K$ with the fibre $\sT^\*_q
Q = \zp_Q^{-1}(q)$ is a vector subspace of the fibre,
    \item {\rm 3)}the Liouville form $\zy_Q$ vanishes on $\sT K$,
        \parindent=0pt\vskip0mm
    then $C \subset Q$ is a submanifold and $K = \sT^\polar C$.
        \endList        }
        \Proof{
    It follows from Proposition~1 that $C \subset Q$ is a submanifold.  We will use certain results established in
the proof of Proposition~1.  The equality \RF(155) implies that
        $$\dim K_q = \dim K - \dim C = \dim Q - \dim C = \dim \sT_q^\polar C
                                                                                                                \eqno(160)$$
    at each $q \in C$.  Let $p$ be an element of $K$ and let $u$ be a vector tangent to $C$ at $q = \zp_Q(p)$.  It follows
from \RF(143) that there exists a vector $v \in \sT_p K$ such that $\sT\zp_Q(v) = u$.  The inclusion
        $$K \subset \sT_q^\polar C
                                                                                                                \eqno(161)$$
    follows from
        $$\langle p, u\rangle = \zy_Q(v) = 0.
                                                                                                                \eqno(162)$$
    The two results \RF(160) and \RF(161) imply that $K = \sT^\polar C$.
        }

        \Definition{
    A {\it cotangent fibration morphism} from the cotangent fibration of a manifold $Q$ to the cotangent fibration of a
manifold $Q'$ is a diagram
    \vskip1mm
        $$\vcenter{\xymatrix@R+3mm @C+10mm{{(\sT^{\*}Q,\zy_Q)} \ar[d]_*{\zp_Q} \ar[r]^*{\zf} &
            (\sT^\* Q',\zy_{Q'}) \ar[d]_*{\zp_{Q'}} \cr
            Q \ar[r]^*{\zc} & Q'}}
                                                                                                                \eqno(163)$$
    \vskip2mm
    \noindent where
    \vskip1mm
        $$\vcenter{\xymatrix@R+3mm @C+10mm{{\sT^{\*}Q} \ar[d]_*{\zp_Q} \ar[r]^*{\zf} &
            \sT^\* Q' \ar[d]_*{\zp_{Q'}} \cr
            Q \ar[r]^*{\zc} & Q'}}
                                                                                                                \eqno(164)$$
    \vskip2mm
    \noindent is a vector fibration morphism, the dimension of $\gr(\zf)$ is equal to the dimension of $Q' \times Q$, and 
        $$(\zy_{Q'} \ominus \zy_Q)|\gr(\zf) = 0.
                                                                                                                \eqno(165)$$
    The form $\zy_{Q'} \ominus \zy_Q$ on $\sT^{\*}Q' \times \sT^{\*}Q$ is defined by
        $$\zy_{Q'} \ominus \zy_Q = pr'{^{\*}}\zy_{Q'} - pr{^{\*}}\zy_Q
                                                                                                                \eqno(166)$$
    with canonical projections
        $$pr \,\colon \sT^{\*}Q' \times \sT^{\*}Q \rightarrow \sT^{\*}Q \;,  \hskip10mm pr' \,\colon \sT^{\*}Q' \times
\sT^{\*}Q \rightarrow \sT^{\*}Q'.
                                                                                                                \eqno(167)$$
        }

        \Definition{
    A {\it Liouville structure morphism} is a diagram
    \vskip1mm
        $$\vcenter{\xymatrix@R+2mm @C+10mm{{(P,\zy)} \ar[d]_*{\zp} \ar[r]^*{\zr} &
            (P',\zy') \ar[d]_*{\zp'} \cr
            Q \ar[r]^*{\zs} & Q'}}
                                                                                                                \eqno(168)$$
    \vskip2mm
    \noindent where
    \vskip1mm
        $$\vcenter{\xymatrix@R+2mm @C+10mm{{(P,\zy)} \ar[d]_*{\zp} \cr Q}}
                \hskip20mm  \;\;{\rm and}\;\; \hskip20mm
        \vcenter{\xymatrix@R+2mm @C+10mm{{(P',\zy')} \ar[d]_*{\zp'} \cr Q}}
                                                                                                                \eqno(169)$$
    \vskip2mm
    \noindent are Liouville structures,
        $$\vcenter{\xymatrix@R+2mm @C+10mm{{P} \ar[d]_*{\zp} \ar[r]^*{\zr} &
            P' \ar[d]_*{\zp'} \cr
            Q \ar[r]^*{\zs} & Q'}}
                                                                                                                \eqno(170)$$
    \vskip2mm
    \noindent is a vector fibration morphism, and one of the following conditions is satisfied.
\vskip1mm       \List
    \item {1)} If the Liouville structures \RF(1) are characterized by isomorphisms
    \vskip1mm
        $$\vcenter{\xymatrix@R+2mm @C+10mm{{P} \ar[d]_*{\zp} \ar[r]^*{\za} &
            \sT^\*Q \ar[d]_*{\zp_Q} \cr
            Q \ar@{=}[r] & Q}}
                                                                                                                \eqno(171)$$
    \vskip0mm
    and
    \vskip-3mm
        $$\vcenter{\xymatrix@R+2mm @C+10mm{{P'} \ar[d]_*{\zp'} \ar[r]^*{\za'} &
            \sT^\*Q' \ar[d]_*{\zp_{Q'}} \cr
            Q' \ar@{=}[r] & Q'}}
                                                                                                                \eqno(172)$$
    \vskip0mm
    then
        $$\vcenter{\xymatrix@R+3mm @C+18mm{{(\sT^{\*}Q,\zy_Q)} \ar[d]_*{\zp_Q} \ar[r]^*{\za' \circ \zr \circ \za^{-1}} &
            (\sT^\* Q',\zy_{Q'}) \ar[d]_*{\zp_{Q'}} \cr
            Q \ar[r]^*{\zs} & Q'}}
                                                                                                                \eqno(173)$$
    \vskip1mm
    is a cotangent fibration morphism.
    \item {2)} If the Liouville structures \RF(169) are characterized by pairings
        $$\langle \;\, ,\;\rangle \,\colon P \fpr{(\zp,\zt_Q)} \sT Q \rightarrow \R
                                                                                                                \eqno(174)$$
    and
        $$\langle \;\, ,\;\rangle' \,\colon P' \fpr{(\zp',\zt_{Q'})} \sT Q' \rightarrow \R,
                                                                                                                \eqno(175)$$
    then the dimension of $\gr(\zr)$ is equal to the dimension of $Q' \times Q$ and
        $$\langle p', v'\rangle - \langle p, v\rangle = 0
                                                                                                                \eqno(176)$$
    for each pair $(p',p) \in \gr(\zr)$ and each pair $(v',v) \in \sT\gr(\zs) \subset \sT Q' \times \sT Q$ such
that $\zt_Q(v) = \zp(p)$ and $\zt_{Q'}(v') = \zp'(p')$.
\dacapo
    \item {3)} If the Liouville structures \RF(169) are characterized by symplectic forms $\zw$ and $\zw'$, then the
relation $\zr$ is a symplectic relation from $(P,\zw)$ to $(P',\zw')$.
    \item {4)} If the Liouville structures \RF(169) are characterized by Liouville forms $\zy$ and $\zy'$, then the
dimension of $\gr(\zr)$ is equal to the dimension of $Q' \times Q$. and
        $$(\zy' \ominus \zy)|\gr(\zr) = 0.
                                                                                                                \eqno(177)$$
    The form $\zy' \ominus \zy$ on $P' \times P$ is defined by
        $$\zy' \ominus \zy = pr'{^{\*}}\zy' - pr{^{\*}}\zy
                                                                                                                \eqno(178)$$
    with canonical projections
        $$pr \,\colon P' \times P \rightarrow P \;,   \hskip10mm pr' \,\colon P' \times P \rightarrow P'.
                                                                                                                \eqno(179)$$
        \endList    }

    We show the equivalence of the listed conditions.
    \vskip1mm
        \List
    \item {A)} Equivalence of 1) and 4).  The Liouville forms $\zy$ and $\zy'$ and the isomorphisms \RF(171) and
\RF(172) characterizing the Liouville structures involved are in the relations
        $$\za^\*\zy_Q = \zy,\;\;\; \za'{}^\*\zy_{Q'} = \zy'.
                                                                                                                \eqno(180)$$
    It follows that
        $$(\zy' \ominus \zy)|\gr(\zr) = 0
                                                                                                                \eqno(181)$$
    if and only if
        $$(\zy_{Q'} \ominus \zy_Q)|\gr(\za' \circ \zr \circ \za^{-1}) = 0.
                                                                                                                \eqno(182)$$
    \item {B)} Equivalence of 2) and 4).  Relations between the Liouville forms $\zy$ and $\zy'$ and the corresponding
pairings \RF(1) and \RF(1) imply the equality
        $$(\zy' \ominus \zy)(w',w) = \zy'(w') - \zy(w) = \langle \zt_{P'}(w'), \sT\zp'(w')\rangle' - \langle \zt_P(w),
\sT\zp(w)\rangle
                                                                                                                \eqno(183)$$
    with $w \in \sT P$ and $w' \in \sT P'$.  It follows from Proposition 1 that 
        $$(\sT\zp' \times \sT\zp)(\sT_{(p',p)}\gr(\zr)) = \sT_{(\zp'(p'),\zp(p))}\gr(\zs)
                                                                                                                \eqno(184)$$
    for each pair $(p',p) \in \gr(\zr)$.  It follows that
        $$(\zy' \ominus \zy)(w',w) = 0
                                                                                                                \eqno(185)$$
    for each pair $(w',w) \in \sT\gr(\zr)$ if and only if
        $$\langle p', v'\rangle - \langle p, v\rangle = 0
                                                                                                                \eqno(186)$$
    for each pair $(p',p) \in \gr(\zr)$ and each pair $(v',v) \in \sT\gr(s)$ such that $\zt_{Q'}(v') = \zp'(p')$ and
$\zt_{Q}(v) = \zp(p)$.
    \item {C)} Equivalence of 3) and 4).  The symplectic forms $\zw$ and $\zw'$ of the condition 3) and the Liouville forms
$\zy$ and $\zy'$ of condition 4) are related by
        $$\zw = \rd\zy,\;\;\;\;  \zw' = \rd\zy'.
                                                                                                                \eqno(187)$$
    Hence,
        $$\zw' \ominus \zw = \rd(\zy' \ominus \zy).
                                                                                                                \eqno(188)$$
    It follows from Proposition 1 that the vector fibration
    \vskip1mm
        $$\vcenter{\xymatrix@R+2mm @C+10mm{{P' \times P} \ar[d]_*{\zp' \times \zp} \cr Q' \times Q}}
                                                                                                                \eqno(189)$$
    \vskip2mm
    induces a vector fibration
    \vskip1mm
        $$\vcenter{\xymatrix@R+2mm @C+10mm{{\gr(\zr)\VPD{6pt}} \ar[d] \cr \gr(\zs)}}
                                                                                                                \eqno(190)$$
    \vskip2mm
    The form
        $$(\zy' \ominus \zy)|\gr(\zr)
                                                                                                                \eqno(191)$$
\vskip0pt
    is linear since $\zy$ and $\zy'$ are linear.  It follows that
        $$(\zw' \ominus \zw)|\gr(\zr) = \rd(\zy' \ominus \zy)|\gr(\zr) = 0
                                                                                                                \eqno(192)$$
\vskip0pt
    if and only if there is a linear function $f$ on $\gr(\zr)$ such that
        $$(\zy' \ominus \zy)|\gr(\zr) = \rd f.
                                                                                                                \eqno(193)$$
\vskip0pt
    The form $\zy' \ominus \zy$ is vertical since $\zy$ and $\zy'$ are vertical.  If the function $f$ exists, then its
\vskip0pt differential $\rd f = (\zy' \ominus \zy)|\gr(\zr)$ is vertical.  It follows that $f = 0$.  We have established the
\vskip0pt equivalence of the equalities
        $$(\zy' \ominus \zy)|\gr(\zr) = 0
                                                                                                                \eqno(194)$$
\vskip0pt
    and
        $$(\zw' \ominus \zw)|\gr(\zr) = 0.
                                                                                                                \eqno(195)$$
\vskip0pt
    The second of these equalities together with the condition that the dimension of $\gr(\zr)$ is equal
    \vskip0pt to the dimension of $Q' \times Q$ imply that the relation $\zr$ is a symplectic relation from $(P,\zw)$ to
\vskip0pt $(P',\zw')$.
        \endList

        \sect{Functors in the category of Liouville structures.}
    Liouville structures and Liouville structure morphisms do not form a category in the usual sense since the composition
of morphisms is not necesarilly a morphism.  We are using the term 'category' in a loose sence.  A true category of
Liouville structures with families of functions acting as morphisms will be presented in a separate publication.  We list
examples of functors.
        \endList

        \ssect{Multiplication by a number.}
    The functor of {\it multiplication by a number} $k \neq 0$ associates the Liouville structure
    \vskip1mm
        $$\vcenter{
        \begindc{0}[1]
        \obj(000,45)[01]{$(P,k\zy)$}
        \obj(000,00)[00]{$Q$}
        \mor{01}{00}[8,8]{$\zp$}[-1,0]
        \enddc}
                                                                                                                \eqno(196)$$
    \vskip2mm
    \noindent with a Lioville structure
    \vskip1mm
        $$\vcenter{
        \begindc{0}[1]
        \obj(000,45)[01]{$(P,\zy)$}
        \obj(000,00)[00]{$Q$}
        \mor{01}{00}[8,8]{$\zp$}[-1,0]
        \enddc}
                                                                                                                \eqno(197)$$
    \vskip2mm
    \noindent and the morphism
    \vskip1mm
        $$\vcenter{\xymatrix@R+2mm @C+10mm{{(P,k\zy)} \ar[d]_*{\zp} \ar[r]^*{\zr} &
            (P',k\zy') \ar[d]_*{\zp'} \cr
            Q \ar[r]^*{\zs} & Q'}}
                                                                                                                \eqno(198)$$
    \vskip2mm
    \noindent with a morphism
    \vskip1mm
        $$\vcenter{\xymatrix@R+2mm @C+8mm{{(P,\zy)} \ar[d]_*{\zp} \ar[r]^*{\zr} &
            (P',\zy') \ar[d]_*{\zp'} \cr
            Q \ar[r]^*{\zs} & Q'}}
                                                                                                                \eqno(199)$$
    \vskip2mm
    The case $k = -1$ is the only important case of this functor.  The case $k = 1$ corresponds to the identity functor.

        \ssect{The direct sum.}
    The {\it direct sum} of two Liouville structures
    \vskip1mm
        $$\vcenter{\xymatrix@R+2mm @C+8mm{{(P_1,\zy_1)} \ar[d]_*{\zp_1} \cr Q_1}}
                \hskip20mm  {\rm and}  \hskip20mm
        \vcenter{\xymatrix@R+2mm @C+10mm{{(P_2,\zy_2)} \ar[d]_*{\zp_2} \cr Q_2}}
                                                                                                                \eqno(200)$$
    \vskip2mm
    \noindent is the Liouville structure

    \vskip1mm
        $$\vcenter{\xymatrix@R+2mm @C+10mm{{(P_2 \times  P_1,\zy_2 \oplus \zy_1)} \ar[d]_*{\zp_2 \times \zp_1} \cr Q_2 \times
Q_1}}
                                                                                                                \eqno(201)$$
    \vskip2mm
    \noindent

        \ssect{The direct difference.}

    The {\it direct difference}
    \vskip1mm
        $$\vcenter{\xymatrix@R+2mm @C+10mm{{(P_2 \times  P_1,\zy_2 \ominus \zy_1)} \ar[d]_*{\zp_2 \times \zp_1} \cr Q_2
\times Q_1}}
                                                                                                                \eqno(202)$$
    \vskip2mm
    \noindent is obtained as the direct sum of Liouville structures
    \vskip1mm
        $$\vcenter{\xymatrix@R+2mm @C+10mm{{(P_1,-\zy_1)} \ar[d]_*{\zp_1} \cr Q_1}}
                \hskip20mm  {\rm and}  \hskip20mm
        \vcenter{\xymatrix@R+2mm @C+10mm{{(P_2,\zy_2)} \ar[d]_*{\zp_2} \cr Q_2}}
                                                                                                                \eqno(203)$$
    \vskip2mm
    \noindent

        \ssect{Derivations $\ri_T$ and $\rd_T$.}
    Let $T \,\colon \sT M \rightarrow \sT M$ be the identity mapping interpreted as a deformation of the tangent projection
$\zt_M \,\colon \sT M \rightarrow M$.  We associate with $T$ linear operators $\ri_T \,\colon \zF(M) \rightarrow \zF(\sT
M)$ and $\rd_T \,\colon \zF(M) \rightarrow \zF(\sT M)$ from the exterior algebra $\zF(M)$ of differential forms on the
manifold $M$ to the exterior algebra $\zF(\sT M)$ of differential forms on $\sT M$.  The constructions of the operators
follow the extension of the Fr\"olicher-Nijenhuis theory [1] presented in [7].  The operator $\ri_T$ is a derivation of
degree -1 relative to the homomorphism $\zt_M{}^\* \,\colon \zF(M) \rightarrow \zF(\sT M)$ in the sense that if $\zm_1$ and
$\zm_2$ are differential forms on $M$ of degrees $r_1$ and $r_2$ respectively, then $\ri_T\zm_1$ and $\ri_T\zm_2$ are
differential forms on $\sT M$ of degrees $r_1 - 1$ and $r_2 - 1$, and the relation
        $$\ri_T(\zm_1 \wedge \zm_2) = \ri_T\zm_1 \wedge \zt_M{}^\*\zm_2 + (-1)^r \zt_M{}^\*\zm_1 \wedge \ri_T\zm_2
                                                                                                                \eqno(204)$$
    holds.  The derivation $\ri_T$ is of type $\ri_\*$ since $\ri_T f = 0$ for each function $f$ on $M$.  The formula
        $$\ri_T\zm(w_1,\ldots , w_r) = \zm(\zt_{\sT M}(w_1), \sT\zt_M(w_1), \ldots, \sT\zt_M(w_r))
                                                                                                                \eqno(205)$$
    gives an explicit construction of the derivation $\ri_T$.  The form $\zm$ is of degree $r + 1$ and $w_1,\ldots,w_r$ are
elements of $\sT\sT M$ such that $\zt_{\sT M}(w_1) = \ldots = \zt_{\sT M}(w_r)$.

    The operator $\rd_T$ is a derivation of degree 0 and type $\rd_\*$.  It is defined by
        $$\rd_T = \ri_T \rd + \rd \ri_T
                                                                                                                \eqno(206)$$
    and is in the relation
        $$\rd_T\rd = \rd\rd_T
                                                                                                                \eqno(207)$$
    with the exterior differentials in $\zF(M)$ and $\zF(\sT M)$.

    The following property of the derivation $\rd_T$ was established in [6].  Given elements
$w_1,w_2,\ldots ,w_r$ of $\sT\sT M$ such that
        $$\zt_{\sT M}(w_1) = \ldots = \zt_{\sT M}(w_r)
                                                                                                                \eqno(208)$$
    it is possible to construct mappings
        $$\zd\zx_1 \,\colon \R \rightarrow \sT M, \ldots , \zd\zx_r \,\colon \R \rightarrow \sT M
                                                                                                                \eqno(209)$$
    such that
        $$w_1 = \zk_M(\st\zd\zx_1(0)), \ldots , w_q = \zk_M(\st\zd\zx_r(0))
                                                                                                                \eqno(210)$$
    and
        $$\zt_M \circ \zd\zx_1 = \cdots = \zt_M \circ \zd\zx_r.
                                                                                                                \eqno(211)$$
    For an $r$-form $\zm \,\colon \PR^r_M \sT M \rightarrow \R$ we have
        $$\rd_T\zm(w_1,w_2,\ldots ,w_r) = \rD(\zm \circ (\zd\zx_1,\zd\zx_2,\ldots ,\zd\zx_r))(0)
                                                                                                                \eqno(212)$$
    with
        $$(\zd\zx_1,\zd\zx_2,\ldots ,\zd\zx_r) \,\colon \R \rightarrow \PR^r_M \sT M \,\colon t \mapsto
(\zd\zx_1(t),\zd\zx_2(t),\ldots ,\zd\zx_r(t)).
                                                                                                                \eqno(213)$$

        \ssect{The tangent functor.}
    The {\it tangent functor} $\sT$ associates the structure
    \vskip1mm
        $$\vcenter{\xymatrix@R+2mm @C+10mm{{(\sT P,\rd_T\zy)} \ar[d]_*{\sT\zp} \cr \sT Q}}
                                                                                                                \eqno(214)$$
    \vskip2mm
    \noindent with a Liouville structure
    \vskip1mm
        $$\vcenter{\xymatrix@R+2mm @C+10mm{{(P,\zy)} \ar[d]_*{\zp} \cr Q}}
                                                                                                                \eqno(215)$$
    \vskip2mm
    \noindent and the morphism
    \vskip1mm
        $$\vcenter{\xymatrix@R+2mm @C+8mm{{(\sT P,\rd_T\zy)} \ar[d]_*{\sT\zp} \ar[r]^*{\sT\zr} &
            (\sT P',\rd_T\zy') \ar[d]_*{\sT\zp'} \cr
            \sT Q \ar[r]^*{\sT\zs} & \sT Q'}}
                                                                                                                \eqno(216)$$
    \vskip2mm
    \noindent with a Liouville structure morphism
    \vskip1mm
        $$\vcenter{\xymatrix@R+2mm @C+12mm{{(P,\zy)} \ar[d]_*{\;\;\;\zp} \ar[r]^*{\zr} &
            (P',\zy') \ar[d]_*{\zp'} \cr
            Q \ar[r]^*{\zs} & Q'}}
                                                                                                                \eqno(217)$$
    \vskip2mm
    \noindent 

    There is an alternative construction of the Liouville structure \RF(214).  The fibrations
    \vskip1mm
        $$\vcenter{\xymatrix@R+2mm @C+10mm{{\sT P} \ar[d]_*{\sT\zp} \cr \sT Q}}
                \hskip20mm  {\rm and}  \hskip20mm
        \vcenter{\xymatrix@R+2mm @C+10mm{{\sT\sT Q} \ar[d]_*{\sT\zt_Q} \cr \sT Q}}
                                                                                                                \eqno(218)$$
    \vskip2mm
    \noindent form a dual pair with a version
        $$\langle \,\;,\;\rangle^{\ssT} \,\colon \sT P \fpr{(\sT\zp,\sT\zt_Q)} \sT\sT Q \rightarrow \R
                                                                                                                \eqno(219)$$
    of the tangent pairig \RF(51) and the fibrations
    \vskip1mm
        $$\vcenter{\xymatrix@R+2mm @C+10mm{{\sT^\*\sT Q} \ar[d]_*{\zp_{\sT Q}} \cr \sT Q}}
                \hskip20mm  {\rm and}  \hskip20mm
        \vcenter{\xymatrix@R+2mm @C+10mm{{\sT\sT Q} \ar[d]_*{\zt_{\sT Q}} \cr \sT Q}}
                                                                                                                \eqno(220)$$
    \vskip2mm
    \noindent are a canonical dual pair with the canonical pairing
        $$\langle \,\;,\;\rangle\VPD{4pt}_{\sT Q} \,\colon \sT^\*\sT Q \fpr{(\zp_{\sT Q},\zt_{\sT Q})} \sT\sT Q \rightarrow
\R. 
                                                                                                                \eqno(221)$$
    The diagram
    \vskip1mm
        $$\vcenter{\xymatrix@R+2mm @C+14mm{{\sT P} \ar[d]_*{\sT\zp} \ar[r]^*{\za_\sT} &
            \sT^\*\sT Q \ar[d]_*{\zp_{\sT Q}} \cr
            \sT Q \ar@{=}[r] & \sT Q}}
                                                                                                                \eqno(222)$$
    \vskip2mm
    \noindent is a vector fibration isomorphism dual to the vector fibration isomorphism
    \vskip1mm
        $$\vcenter{\xymatrix@R+2mm @C+14mm{{\sT\sT Q} \ar[d]_*{\;\;\sT\zt_Q} &
            \sT\sT Q \ar[l]_*{\zk_Q} \ar[d]_*{\zt_{\sT Q}} \cr
            \sT Q \ar@{=}[r] & \sT Q}}
                                                                                                                \eqno(223)$$
    \vskip2mm
    \noindent extracted from the double vector fibration isomorphism \RF(47).  The duality is expressed by
        $$\langle \za_\sT(\dot p), \zd\dot q\rangle\VPD{4pt}_{\sT Q} = \langle \dot p, \zk_Q(\zd\dot q)\rangle^{\ssT}.
                                                                                                                \eqno(224)$$

    The equality
        $$\za_\sT^\*\zy_{\sT Q} = \rd_T\zy
                                                                                                                \eqno(225)$$
    shows that the diagrams \RF(214) and \RF(222) represent the same Liouville structure.  The following objects will be
used in the proof of this equality.  With an element $\zd\dot p$ of $\sT\sT P$ we associate a curve
        $$\zd\zh \,\colon \R \rightarrow \sT P
                                                                                                                \eqno(226)$$
    such that
        $$\zd\dot p = \zk_P(\st\zd\zh(0)).
                                                                                                                \eqno(227)$$
    We introduce curves
        $$\zh \,\colon \R \rightarrow P,\;\;\;\dot\zh \,\colon \R \rightarrow \sT P,\;\;\;\zx \,\colon \R \rightarrow
Q,\;\;\;\zd\zx \,\colon \R \rightarrow \sT Q,\;\;\;\zd\dot\zh \,\colon \R \rightarrow \sT\sT P,\;\;\;\zd\dot\zx \,\colon \R
\rightarrow \sT\sT Q
                                                                                                                \eqno(228)$$
    defined by
        $$\zh = \zt_P \circ \zd\zh,\;\;\;\dot\zh = \st\zh,\;\;\;\zx = \zp \circ \zh,\;\;\;\zd\zx = \sT\zp \circ
\zd\zh,\;\;\;\zd\dot\zh = \zk_P \circ \st\zd\zh,\;\;\;\zd\dot\zx = \sT\sT\zp \circ \zd\dot\zh.
                                                                                                                \eqno(229)$$
    The equality \RF(225) follows from
        $$\eqalign{
    (\za_\sT^\*\zy_{\sT Q})(\zd\dot p)  =& \zy_{\sT Q}(\sT\za_\sT(\zd\dot p)) \cr
            &= \langle \zt_{\sT\*\sT Q}(\sT\za_\sT(\zd\dot p)), \sT\zp_{\sT Q}(\sT\za_\sT(\zd\dot p)) \rangle\VPD{4pt}_{\sT
Q} \cr
            &= \langle \za_\sT(\zt_{\sT P}(\zd\dot p)), \sT(\zp_{\sT Q} \circ \za_\sT)(\zd\dot p)) \rangle\VPD{4pt}_{\sT Q}
\cr
            &= \langle \za_\sT(\zt_{\sT P}(\zd\dot p)), \sT\sT\zp(\zd\dot p)) \rangle\VPD{4pt}_{\sT Q} \cr
            &= \langle \za_\sT(\dot p), \zd\dot q \rangle\VPD{4pt}_{\sT Q} \cr
            &= \langle \dot p, \zk_Q(\zd\dot q) \rangle^{\ssT} \cr
            &= \rD\langle \zh, \zd\zx\rangle(0) }
                                                                                                                \eqno(230)$$
    and
        $$\rd_T\zy(\zd\dot p) = \rD(\zy \circ \zd\zh)(0) = \rD\langle \zh, \zd\zx\rangle(0),
                                                                                                                \eqno(231)$$
    where we have used the definition \RF(224) of $\za_\sT$, the formula \RF(212) applied to $\rd_T\zy$, the relation
\RF(102), and definitions
        $$\dot p = \zt_{\sT P}(\zd\dot p) = \dot\zh(0)
                                                                                                                \eqno(232)$$
    and
        $$\zd\dot q = \sT\sT\zp(\zd\dot p) = \zd\dot\zx(0) = \zk_Q(\st\zd\zx(0)).
                                                                                                                \eqno(233)$$

    Two more definitions of the tangent Liouville structure are possible.  The pairing
        $$\langle \,\;, \;\rangle \,\colon \sT P \fpr{(\sT\zp,\zt_{\sT Q})} \sT\sT Q \rightarrow \R \,\colon (\dot
p,\zd\dot q) \mapsto \langle \dot p, \zk_Q(\zd\dot q)\rangle^{\ssT} = \langle \za_\sT(\dot p), \zd\dot
q\rangle\VPD{4pt}_{\sT Q}
                                                                                                                \eqno(234)$$
    or the linear symplectic form $\rd\rd_T\zy$ on the vector fibration
    \vskip1mm
        $$\vcenter{\xymatrix@R+2mm @C+10mm{{\sT P} \ar[d]_*{\sT\zp} \cr \sT Q}}
                                                                                                                \eqno(235)$$
    \vskip2mm
    \noindent can be used.

    The tangent structure \RF(214) is a Liouville structure for the symplectic structure $(\sT P,\rd_T\rd\zy)$.

        \ssect{The Hamilton functor.}

    The {\it Hamilton functor} $\sH$ is a covariant functor from the category of symplectic manifolds to the category of
Liouville structures.  It associates the Liouville structure
    \vskip1mm
        $$\vcenter{\xymatrix@R+2mm @C+10mm{{(\sT P,\ri_T\zw)} \ar[d]_*{\zt_P} \cr P}}
                                                                                                                \eqno(236)$$
    \vskip2mm
    \noindent with a symplectic manifold $(P,\zw)$, and the morphism
    \vskip1mm
        $$\vcenter{\xymatrix@R+2mm @C+10mm{{(\sT P,\ri_T\zw)} \ar[d]_*{\sT\zp} \ar[r]^*{\sT\zf} &
            (\sT P',\ri_T\zw') \ar[d]_*{\zt_P'} \cr
            P \ar[r]^*{\zf} & P'}}
                                                                                                                \eqno(237)$$
    \vskip2mm
    \noindent with a symplectomorphism $\zf \,\colon (P,\zw) \rightarrow (P',\zw')$.

    The formula
        $$\ri_T\zw(\zd\dot p) = \zw(\dot p,\zd p)
                                                                                                                \eqno(238)$$
    with $\dot p = \zt_{\sT P}(\zd\dot p)$ and $\zd p = \sT\zt_P(\zd\dot p)$ defines the Liouville form $\ri_T\zw$.

    The diagram
    \vskip1mm
        $$\vcenter{\xymatrix@R+2mm @C+14mm{{\sT P} \ar[d]_*{\sT\zp} \ar[r]^*{\zb_{(P.\zw)}} &
            \sT^\*\sT Q \ar[d]_*{\zp_P} \cr
            P \ar@{=}[r] & P}}
                                                                                                                \eqno(239)$$
    \vskip2mm
    \noindent with the mapping $\zb_{(P.\zw)}$ characterized by
        $$\langle \zb_{(P.\zw)}(\dot p), \zd p\rangle\VPD{4pt}_P = \zw(\dot p,\zd p)
                                                                                                                \eqno(240)$$
    defines a Liouville structure equivalent to \RF(236).  The equality
        $$\zb_{(P.\zw)}{}^\* \zy_P = \ri_T\zw
                                                                                                                \eqno(241)$$
    establishes the equivalence.  This equality is proved by
        $$\eqalign{
    (\zb_{(P.\zw)}{}^\* \zy_P)(\zd\dot p)  =& \zy_P(\sT\zb_{(P.\zw)}(\zd\dot p)) \cr
            &= \langle \zt_{\sT\*P}(\sT\zb_{(P.\zw)}(\zd\dot p)), \sT\zp_P(\sT\zb_{(P.\zw)}(\zd\dot p))
\rangle\VPD{4pt}_{\sT Q} \cr
            &= \langle \zb_{(P.\zw)}(\zt_{\sT P}(\zd\dot p)), \sT(\zp_P \circ \zb_{(P.\zw)})(\zd\dot p)
\rangle\VPD{4pt}_{\sT Q} \cr
            &= \langle \zb_{(P.\zw)}(\zt_{\sT P}(\zd\dot p)), \sT\zt_P(\zd\dot p) \rangle\VPD{4pt}_{\sT Q} \cr
            &= \langle \zb_{(P.\zw)}(\dot p), \zd p \rangle\VPD{4pt}_{\sT Q} \cr
            &= \zw(\dot p, \zd p) \cr
            &= \ri_T\zw(\zd\dot p). }
                                                                                                                \eqno(242)$$
    We are using the conventions introduced in the preceding section.

    The pairing
        $$\langle \,\;, \;\rangle \,\colon \sT P \fpr{(\zt_P,\zt_P)} \sT P \rightarrow \R \,\colon (\dot p,\zd p)
\mapsto \zw(\dot p,\zd p)
                                                                                                                \eqno(243)$$
    provides an alternative definition of the Hamilton Liouville structure.

    If the Hamilton functor is applied to the symplectic $(P,\rd\zy)$ associated with the Liouville structure \RF(215), then
a second Liouville structure for the symplectic structure $(\sT P,\rd_T\rd\zy)$ is obtained.  The diagram
    \vskip1mm
        $$\vcenter{
        \begindc{0}[1]
        \obj(000,90)[02]{$\sT^\*P$}   \obj(120,90)[12]{$\sT P$}   \obj(240,90)[22]{$\sT^\*\sT Q$}
                    \obj(60,45)[01]{$P$}   \obj(180,45)[11]{$\sT Q$}
                            \obj(120,00)[00]{$Q$}
        \mor{12}{02}[12,14]{$\zb_{(P,\zw)}$}[-1,0]  \mor{12}{22}[12,16]{$\za_\sT$}[1,0]
    \mor{02}{01}[10,8]{$\zp_P$}[-1,0]   \mor{12}{01}[10,8]{$\zt_P$}[1,0]
                    \mor{12}{11}[10,12]{$\sT\zp$}[-1,0] \mor{22}{11}[10,11]{$\zp_{\sT Q}$}[1,0]
        \mor{01}{00}[10,8]{$\zp$}[-1,0] \mor{11}{00}[10,8]{$\zt_Q$}[1,0]
        \enddc}
                                                                                                                \eqno(244)$$
    \vskip2mm
    \noindent collects together some of the structures present in this case.

        \ssect{The phase functor.}

    The {\it cotangent functor} is a contravariant functor from the category of differential manifolds to the category of
Liouville structures.  It associates the cotangent Liouville structure
    \vskip1mm
        $$\vcenter{\xymatrix@R+2mm @C+10mm{{(\sT^\*Q,\zy_Q)} \ar[d]_*{\zp{}_Q} \cr Q}}
                                                                                                                \eqno(245)$$
    \vskip2mm
    \noindent with a differential manifold $Q$ and the morphism
    \vskip1mm
        $$\vcenter{\xymatrix@R+2mm @C+10mm{{(\sT^\*Q',\zy_{Q'})} \ar[d]_*{\zp_{Q'}} \ar[r]^*{\sT^\*\zc} &
            (\sT^\*Q,\zy_Q) \ar[d]_*{\zp_Q} \cr
            Q' \ar[r]^*{\zc^{-1}} & Q}}
                                                                                                                \eqno(246)$$
    \vskip2mm
    \noindent with a diffeomorphism or a reduction $\zc \,\colon Q \rightarrow Q'$.

    The {\it phase functor} is a covariant version of the cotangent functor.  It associates the cotangent Liouville
structure \RF(245) with a differential manifold $Q$ and the morphism
    \vskip1mm
        $$\vcenter{\xymatrix@R+2mm @C+10mm{{(\sT^\*Q,\zy_Q)} \ar[d]_*{\zp_Q} \ar[r]^*{\sT^\*\zc^{-1}} &
            (\sT Q',\zy_{Q'}) \ar[d]_*{\zp_{Q'}} \cr
            Q \ar[r]^*{\zc} & Q'}}
                                                                                                                \eqno(247)$$
    \vskip2mm
    \noindent with a diffeomorphism or a reduction $\zc \,\colon Q \rightarrow Q'$.

        \sect{Generating functions.}
    A Liouville structure offers the possibility of generating from {\it generating objects} subsets of a symplectic
manifold $(P,\zw)$ for which the Liouville structure is established.  Such subsets are usually Lagrangian submanifolds.  Let
    \vskip1mm
        $$\vcenter{\xymatrix@R+2mm @C+10mm{{P} \ar[d]_*{\zp} \ar[r]^*{\za} &
            \sT^\*Q \ar[d]_*{\zp_Q} \cr
            Q \ar@{=}[r] & Q}}
                                                                                                                \eqno(248)$$
    \vskip2mm
    \noindent be a Liouville structure for the symplectic manifold $(P,\zw)$.  The simplest example of a generating object
is a {\it generating function}
        $$U \,\colon Q \rightarrow \R.
                                                                                                                \eqno(249)$$
    The set
        $$S = \za^{-1}(\im(\rd U)) = \left\{f \in P \;;\; \all{\zd q \in \sT Q} \;\;{\rm if}\;\; \zt_Q(\zd q) = \zp(f),
\;\;{\rm then}\;\; \langle \za(f), \zd q\rangle\VPD{4pt}_Q = \langle \rd U, \zd q\rangle\VPD{4pt}_Q \right\}
                                                                                                                \eqno(250)$$
    is a Lagrangian submanifold of $(P,\zw)$ generated by the function $U$.  If the Liouville structure is characterized by
the pairing
        $$\langle \;\, ,\;\rangle \,\colon P \fpr{(\zp,\zt_Q)} \sT Q \rightarrow \R,
                                                                                                                \eqno(251)$$
    then
        $$S = \left\{f \in P \;;\; \all{\zd q \in \sT Q} \;\;{\rm if}\;\; \zt_Q(\zd q) = \zp(f), \;\;{\rm then}\;\; \langle
f, \zd q\rangle = \langle \rd U, \zd q\rangle\VPD{3pt}_Q \right\}
                                                                                                                \eqno(252)$$
    is the set generated by the function $U$.

    A more general example of a generating object is a {\it constrained generating function}
        $$U \,\colon C \rightarrow \R,
                                                                                                                \eqno(253)$$
    defined on a submanifold of $C \subset Q$.  The set
        $$S = \left\{f \in P \;;\; \zp_Q(f)\in C  \;\;{\rm and}\;\; \all{\zd q \in \sT C} \;\;{\rm if}\;\; \zt_Q(\zd q) = \zp(f),
\;\;{\rm then}\;\; \langle f, \zd q\rangle = \langle \rd U, \zd q\rangle\VPD{4pt}_C \right\}
                                                                                                                \eqno(254)$$
    is the Lagrangian submanifold of $(P,\zw)$ generated by the constrained function $U$.  This submanifold is an affine
bundle over $C$, modelled on the vector bundle $\sT^\polar C$.

    If $S$ is a Lagrangian submanifold of $(P,\zw)$, then $\zw|S =0$.  Hence, the form $\zy|S$ is closed.  Assuming that
this form is exact we choose a function $\wU \,\colon S \rightarrow \R$ such that $\zy|S = \rd \wU$.  This function is
called a {\it proper function} of $S$. Since the form $\zy$ is vertical, $\langle \rd \wU, \zd q\rangle = 0$ for each $\zd
q \in \sT S\cap \sV P$. This implies that $\wU$ is constant on connected submanifolds in $S\cap P_q$.  If $C = \zp(S)$ is a
submanifold of $Q$ and the fibration $\zp$ restricted to $S$ induces an affine fibration $\zh \,\colon S \rightarrow C$,
then $\wU$ is constant on fibres of $\zh$ and $S$ is generated by a constrained function $U \,\colon C \rightarrow \R$ such
that $U \circ \zh = \wU$. 

\vskip5mm

        \sect{Examples.}
        \Example{
    Let $Q$ be the {\it configuration space} of a static mechanical system and let $U \,\colon Q \rightarrow \R$ be the
internal energy of the system.  The function $U$ generates the {\it constitutive set}
        $$S = \im(\rd U) = \left\{f \in \sT^\*Q \;;\; \all{\zd q \in \sT Q} \;\;{\rm if}\;\; \zt_Q(\zd q) = \zp_Q(f),
\;\;{\rm then}\;\; \langle f, \zd q\rangle\VPD{3pt}_Q = \langle \rd U, \zd q\rangle\VPD{3pt}_Q \right\}.
                                                                                                                \eqno(255)$$
    The Liouville structure in this example is the canonical Liouville structure
    \vskip1mm
        $$\vcenter{\xymatrix@R+2mm @C+10mm{{(\sT^\*Q,\zy_Q)} \ar[d]_*{\zp{}_Q} \cr Q}}
                                                                                                                \eqno(256)$$
    \vskip2mm
        }

        \Example{
    Let $Q$ be the {\it configuration space} of an autonomous mechanical system and let
        $$W_{[t_0,t_1]} \,\colon Q \times Q \rightarrow \R
                                                                                                                \eqno(257)$$
     be the principal Hamilton function for the time interval $[t_0,t_1] \subset \R$.  The set
        $$\eqalign{
            \hskip-10mm D_{[t_0,t_1]} &= \left\{(p_1,p_0) \in \sT^\*Q \times \sT^\*Q\;;\; \all{(\zd q_1,\zd q_0) \in \sT Q
\times \sT Q}
        \;\;{\rm if}\;\; \zt_Q(\zd q_1) = \zp_Q(p_1) \right. \cr
        &\hskip2mm \left.\VPU{11pt} \;\;{\rm and}\;\; \zt_Q(\zd q_0) = \zp_Q(p_0), \;\;{\rm then}\;\; \langle p_1, \zd
q_1\rangle\VPD{4pt}_Q - \langle p_0, \zd q_0\rangle\VPD{4pt}_Q = \langle \rd W_{[t_0,t_1]}, (\zd q_1,\zd
q_0)\rangle\VPD{4pt}_Q \right\}}
                                                                                                                \eqno(258)$$
    is the dynamics in the interval $[t_0,t_1]$ without external forces.  The Liouville structure
    \vskip1mm
        $$\vcenter{\xymatrix@R+2mm @C+10mm{{(\sT^\*Q \times  \sT^\*Q,\zy_Q \ominus \zy_Q)} \ar[d]_*{\zp_Q \times \zp_Q} \cr Q
\times Q}}
                                                                                                                \eqno(259)$$
    \vskip2mm
    \noindent is used.
        }

        \Example{
    Let $Q$ be the {\it configuration space} of an autonomous mechanical system and let
        $$L \,\colon \sT Q \rightarrow \R
                                                                                                                \eqno(260)$$
     be the Lagrangian.  The set
        $$D = \left\{\dot p \in \sT\sT^\*Q\;;\; \all{\zd\dot q \in \sT\sT Q}
        \;\;{\rm if}\;\; \zt_{\sT Q}(\zd\dot q) = \sT\zp_Q(\dot p), \;\;{\rm then}\;\; \langle \za_Q(\dot p), \zd\dot
q\rangle\VPD{4pt}_{\sT Q} = \rd L(\zd\dot q) \right\}
                                                                                                                \eqno(261)$$
    is the Lagrangian representation of infinitesimal dynamics.  The infinitesimal dynamics is an implicit differential
equation for phase space trajectories of the system.  The Liouville structure used is the tangent Liouville structure
derived from the canonical structure \RF(256).  It is represented by the diagram
    \vskip1mm
        $$\vcenter{\xymatrix@R+2mm @C+10mm{{(\sT\sT^\*Q,\rd_T\zy_Q)} \ar[d]_*{\sT\zp_Q} \cr \sT Q}}
                                                                                                                \eqno(262)$$
    \vskip2mm
    \noindent or by the morphism
    \vskip1mm
        $$\vcenter{\xymatrix@R+2mm @C+14mm{{\sT\sT\* Q} \ar[d]_*{\sT\zp_Q} \ar[r]^*{\za_Q} &
            \sT^\*\sT Q \ar[d]_*{\zp_{\sT Q}} \cr
            \sT Q \ar@{=}[r] & \sT Q}}
                                                                                                                \eqno(263)$$
        }

        \Example{
    Let $(P,\zw)$ be the symplectic {\it phase space} of an autonomous mechanical system and let
        $$H \,\colon P \rightarrow \R
                                                                                                                \eqno(264)$$
     be the Hamiltonian.  The set
        $$D = \left\{\dot p \in \sT P\;;\; \all{\zd p \in \sT P} \;\;{\rm if}\;\; \zt_P(\zd p) = \zt_P(\dot p), \;\;{\rm
then}\;\; \zw(\dot p,\zd p) = - \rd H(\zd p) \right\}
                                                                                                                \eqno(265)$$
    is the Hamiltonian  dynamics of the system.  The Liouville structure used is the Hamilton Liouville structure
represented by the pairing \RF(243) and $-H$ is the generating function.  The dynamics $D$ is a differential equation for
phase space trajectories of the system.  It is the image of the Hamiltonian vector field $X \,\colon P \rightarrow \sT P$
defined by
        $$\ri_X \zw = - \rd H.
                                                                                                                \eqno(266)$$
            }

        \Example{
    Let
    \vskip1mm
        $$\vcenter{\xymatrix@R+2mm @C+10mm{{(P_1,\zy_1)} \ar[d]_*{\zp_1} \cr Q_1}}
                \hskip20mm   \hskip20mm
        \vcenter{\xymatrix@R+2mm @C+10mm{{(P_2,\zy_2)} \ar[d]_*{\zp_2} \cr Q_2}}
                                                                                                                \eqno(267)$$
    \vskip2mm
    \noindent be Liouville structures and let
        $$\langle \,\;, \;\rangle_1 \,\colon P_1 \times \sT Q_1  \rightarrow \R
                                                                                                                \eqno(268)$$
    and
        $$\langle \,\;, \;\rangle_2 \,\colon P_2 \times \sT Q_2  \rightarrow \R
                                                                                                                \eqno(269)$$
    be the pairings belonging to these structures.  A differentiable function
        $$F \,\colon Q_2 \times Q_1 \rightarrow \R
                                                                                                                \eqno(270)$$
    generates the graph
        $$\eqalign{
            G &= \left\{(p_2,p_1) \in P_2 \times P_1\;;\; \all{(\zd q_2,\zd q_1) \in \sT Q_2 \times \sT Q_1} \;\;{\rm
if}\;\; \zt_{Q_1}(\zd q_1) = \zp_1(p_2) \right. \cr
        &\hskip10mm\left.{\rm and}\;\;\zt_{Q_2}(\zd q_2) = \zp_2(p_2), \;\;{\rm then}\;\; \langle p_2, \zd q_2\rangle_2 -
\langle p_1, \zd q_1\rangle_1 = \rd F (\zd q_2,\zd q_1) \VPD{5pt}\right\}}
                                                                                                                \eqno(271)$$
    of a symplectic relation
        $$\zg \,\colon (P_1,\zw_1) \rightarrow (P_2,\zw_2).
                                                                                                                \eqno(272)$$
    We are using the pairing
        $$\langle \,\;, \;\rangle \,\colon ((P_2 \times P_1) \times (\sT Q_2 \times \sT Q_1)) \rightarrow \R \,\colon
((p_2,p_1),(\zd q_2,\zd q_1)) \mapsto \langle p_2, \zd q_2\rangle_2 - \langle p_1, \zd q_1\rangle_1
                                                                                                                \eqno(273)$$
    belonging to the Liouville structure
    \vskip1mm
        $$\vcenter{\xymatrix@R+2mm @C+10mm{{(P_2 \times  P_1,\zy_2 \ominus \zy_1)} \ar[d]_*{\zp_2 \times \zp_1} \cr Q_2
\times Q_1}}
                                                                                                                \eqno(274)$$
    \vskip2mm
    \noindent 
        }

        \Example{
    Let
    \vskip1mm
        $$\vcenter{\xymatrix@R+2mm @C+10mm{{(\sT P,\rd_T\zy)} \ar[d]_*{\sT\zp} \cr \sT Q}}
                \hskip40mm
        \vcenter{\xymatrix@R+2mm @C+10mm{{(\sT P,\ri_T\rd\zy)} \ar[d]_*{\zt_P} \cr P}}
                                                                                                                \eqno(275)$$
    \vskip2mm
    \noindent be Liouville structures derived from a Liouville structure
    \vskip1mm
        $$\vcenter{\xymatrix@R+2mm @C+10mm{{(P,\zy)} \ar[d]_*{\zp} \cr Q}}
                                                                                                                \eqno(276)$$
    \vskip2mm
\dacapo
    The pairing
        $$\langle \,\; ,\; \rangle\VPD{4pt} \,\colon P \fpr{(\zp,\zt_Q)} \sT Q \rightarrow \R
                                                                                                                \eqno(277)$$
    is a differentiable function defined on the submanifold $\VPD{10pt}P \fpr{(\zp,\zt_Q)} \sT Q \subset P \times \sT
Q$ and the diagonal $\zD$ of $\sT P \times \sT P$ is the graph of the identity symplectomorphism in $(\sT P,\rd_T\zw)$.  We
will show that the diagonal $\zD$ is generated by the function $-\,\langle \,\; ,\; \rangle$.
        \dacapo
    Evaluation of the form $\ri_T\rd\zy \ominus \rd_T\zy$ on an element $(\dot p,\dot p) \in \zD$ produces the result
        $$(\ri_T\rd\zy \ominus \rd_T\zy)(\dot p,\dot p) = \ri_T\rd\zy(\dot p) - \rd_T\zy(\dot p) = -\rd\ri_T\zy(\dot p).
                                                                                                                \eqno(278)$$
    It follows that the form $\ri_T\rd\zy \ominus \rd_T\zy$ restricted to $\zD$ is the differential of the function
        $$F \,\colon \zD \rightarrow \R \,\colon (\dot p,\dot p) \mapsto -\ri_T\zy(\dot p) = -\langle \zt_P(\dot
p),\sT\zp(\dot p)\rangle
                                                                                                                \eqno(279)$$
    The function $F$ is the proper function of $\zD$.
        \dacapo
    The mapping $\zt_P \times \sT\zp \,\colon \sT P \times \sT P \rightarrow P \times \sT Q$ projects the diagonal $\zD$
onto the submanifold $\VPD{10pt}P \fpr{(\zp,\zt_Q)} \sT Q \subset P \times \sT Q$.  The mapping
        $$\zh \,\colon \zD \rightarrow P \fpr{(\zp,\zt_Q)} \sT Q \,\colon (\dot p,\dot p) \mapsto (\zt_P(\dot p),\sT\zp(\dot p))
                                                                                                                \eqno(280)$$
    is the induced mapping.  The pull back $\langle \,\; ,\; \rangle \circ \zh$ of the pairing to $\zD$ is the function
        $$\langle \,\; ,\; \rangle \circ \zh \,\colon \zD \rightarrow \R \,\colon (\dot p,\dot p) \mapsto \langle
\zt_P(\dot p),\sT\zp(\dot p)\rangle.
                                                                                                                \eqno(281)$$
    Hence, $-\langle \,\; ,\; \rangle \circ \zh = F$.
            \dacapo
    We will show that the mapping $\zh$ is an affine fibration.  We will adopt the obvious identification of $\zD$ with
$\sT P$.  The mapping $\zh$ is identified with
        $$\w\zh \,\colon \sT P \rightarrow P \fpr{(\zp,\zt_Q)} \sT Q \,\colon \dot p \mapsto (\zt_P(\dot p),\sT\zp(\dot p)).
                                                                                                                \eqno(282)$$
    Let $\dot p$ and $\dot p'$ be elements of $\sT P$ such that $\w\zh(\dot p') = \w\zh(\dot p) = (p,\dot q)$.  The
equality $\zt_P(\dot p') = \zt_P(\dot p) = p$ implies that $\zt_P(\dot p' - \dot p) = p$ and from $\sT\zp(\dot p') =
\sT\zp(\dot p) = \dot q$ it follows that $\sT\zp(\dot p' - \dot p) = O_{\zt_Q}(\zt_Q(\dot q))$.  Consequently, $\dot p' -
\dot p \in \sV_p P$.  We see that the mapping $\w\zh$, hence $\zh$, is an affine fibration with the vector fibration
        $$\sV P \fpr{(\zp \circ \zu_P,\zt_Q)} \sT Q \rightarrow P \fpr{(\zp,\zt_Q)} \sT Q \,\colon (p',\dot q) \mapsto
(\zt_P(p'),\dot q)
                                                                                                                \eqno(283)$$
    as its model.  The mapping $\zu_P \,\colon \sV P \rightarrow P$ is the restriction of $\zt_P$ to $\sV P$.  From $\dot p
\in \sT_p P$ and $p' \sV_p P$ we construct the vector  $(p',\sT\zp(\dot p)) \in \sV P \fpr{(\zp \circ \zu_P,\zt_Q)} \sT Q$.
The sum $\dot p + (p',\sT\zp(\dot p))$ is defined as $\dot p + p'$.
        \dacapo
    We have completed the proof that the diagonal $\zD$ is generated by the function $-\,\langle \,\; ,\; \rangle$.
            }

 \sect{References.}

    \item{[1]} A. Fr\"olicher and A. Nijenhuis, {\it Theory of vector valued differential forms}, Nederl. Akad. Wetensch.
Proc. A {\bf 59} (1056), 338--359.

    \item{[2]} J. Grabowski and M. Rotkiewicz, {\it Higher vector bundles and multi-graded symplectic manifolds}, arXiv:marh/070277v1.

    \item{[3]} K.~Konieczna and P.~Urba\'nski, {\it Double vector
bundles and duality}, Arch. Math. (Brno) {\bf 35}, (1999), 59--95.

    \item{[4]} P. Liebermann, {\it On Liouville Forms} in: Poisson Geometry (J. Grabowski and P. Urba\'nski eds.) BCP {\bf 51}, Warsaw 2000, 151--164.   
    
    \item{[5]} K. C. H.  Mackenzie {\it Double Lie algebroids and second-order geometry I}, Adv. Math. {\bf 94}, (1992), 180--239.

    \item{[6]} G. Marmo, W. Tulczyjew and  P. Urba\'nski, {\it Dynamics of autonomous systems with external forces},  Acta Phys. Polon. B {\bf 33} (2002), 1181--1240.

    \item{[7]} G. Pidello and W. M. Tulczyjew, {\it Derivations of differential forms on jet bundles}, Annali di
Matematica pura ed applicata (IV) CXLVII (1987), 249--265.

    \item{[8]} J. Pradines, {\it Fibr\'es vectoriels doubles et calcul des jets non holonomes}, Notes polycopi\'ees, Amiens, 1974.
 
   \item{[9]}W. M. Tulczyjew, {\it Hamiltonian Systems, Lagrangian Systems and the Legendre Transformation}, Symposia Mathematica, {\bf 16} (1974), 247-258.  
   
    \item{[10]}  W. M. Tulczyjew and P. \ Urba\'nski, {\it Differential forms on vector bundles}, Rep. Math. Phys. {\bf 45} (2000), 357--370. 
    
   \end

QQQQQQQQQQQQ To jest stara wersia: QQQQQQQQQQQQQQQQQQQQ

    If $S$ is a Lagrangian submanifold of $(P,\zw)$, then $\zw|S =0$.  Hence, the form $\zy|S$ is closed.  Assuming that
this form is exact we choose a function $\wU \,\colon S \rightarrow \R$ such that $\zy|S = \rd \wU$.  This function is
called a {\it proper function} of $S$.  In the simple case when $S$ is the image of a section $\zs \,\colon Q \rightarrow
P$ of $\zp$ the function $U = \wU \circ \zs$ is a generating function of $S$.  If $C = \zp(S)$ is a submanifold of $Q$ and
the fibration $\zp$ restricted to $S$ induces an affine fibration $\zh \,\colon S \rightarrow C$, then $S$ is generated by
a constrained function $U \,\colon C \rightarrow \R$ such that $U \circ \zh = \wU$. !!!!!!!!!  Prove that the proper
function is constant of fibres of $\zh$.  Hence $U$ is well defined.  Prove that $U$ generates $S$.  !!!!!!!!!!!!

CCCCCCCCCC To ma byc poprawiona wersia: CCCCCCCCCCCCCCCCCC
